%% file: manuscript_stick-breaking-v19_rk_17.01.2018-arxiv.tex
\documentclass[headings=standardclasses]{scrartcl}

\usepackage{geometry}
\usepackage[utf8]{inputenc}
\usepackage[T1]{fontenc}
\usepackage{lmodern}
\usepackage[singlespacing]{setspace}
\usepackage{microtype}
\usepackage{amsmath}
\usepackage{amssymb}
\usepackage{amsthm}
\usepackage{tabulary}
\usepackage{longtable}
\usepackage{booktabs} 
\usepackage{ragged2e}
\newcommand{\ra}[1]{\renewcommand{\arraystretch}{#1}}
\usepackage{siunitx}
\usepackage{bm}
\usepackage[pdftex]{graphicx}
\usepackage{rotating}
\usepackage{pdfpages}
\usepackage{color}
\usepackage{array}
\usepackage[font=bf]{caption}
\usepackage{subcaption}
\usepackage{pgfplots}
\usepackage{pgfplotstable}
\usepackage{float}
\usepackage{appendix}
\pgfplotsset{compat=1.9}
\usepackage{url}
\usepackage[section]{placeins}
\usepackage{natbib}

\usepackage{tikz}
\usetikzlibrary{bayesnet}

\begin{document}
\begin{spacing}{1.2}
\begin{flushleft}
\huge \textbf{A Dirichlet Process Mixture Model of Discrete Choice} \\
\bigskip 
\normalsize
19 January, 2018 \\
\vspace{\baselineskip}
Rico Krueger (corresponding author)\\
Research Centre for Integrated Transport Innovation, School of Civil and Environmental Engineering, UNSW Australia, Sydney NSW 2052, Australia \\
r.krueger@student.unsw.edu.au \\
\vspace{\baselineskip}
Akshay Vij \\
Institute for Choice, University of South Australia\\
140 Arthur Street, North Sydney NSW 2060, Australia\\
vij.akshay@gmail.com \\
\vspace{\baselineskip}
Taha H. Rashidi \\
Research Centre for Integrated Transport Innovation, School of Civil and Environmental Engineering, UNSW Australia, Sydney NSW 2052, Australia\\
rashidi@unsw.edu.au\\
\end{flushleft}
\end{spacing}

\newpage
\section*{Abstract}

We present a mixed multinomial logit (MNL) model, which leverages the truncated stick-breaking process representation of the Dirichlet process as a flexible nonparametric mixing distribution. The proposed model is a Dirichlet process mixture model and accommodates discrete representations of heterogeneity, like a latent class MNL model. Yet, unlike a latent class MNL model, the proposed discrete choice model does not require the analyst to fix the number of mixture components prior to estimation, as the complexity of the discrete mixing distribution is inferred from the evidence. For posterior inference in the proposed Dirichlet process mixture model of discrete choice, we derive an expectation maximisation algorithm. In a simulation study, we demonstrate that the proposed model framework can flexibly capture differently-shaped taste parameter distributions. Furthermore, we empirically validate the model framework in a case study on motorists' route choice preferences and find that the proposed Dirichlet process mixture model of discrete choice outperforms a latent class MNL model and mixed MNL models with common parametric mixing distributions in terms of both in-sample fit and out-of-sample predictive ability. Compared to extant modelling approaches, the proposed discrete choice model substantially abbreviates specification searches, as it relies on less restrictive parametric assumptions and does not require the analyst to specify the complexity of the discrete mixing distribution prior to estimation.

\newpage

\section{Introduction} \label{S_intro}

\subsection{Background and motivation}

The representation of inter-individual taste heterogeneity is a key concern of discrete choice analysis, as information on the distribution of tastes is critical for demand forecasting, welfare analysis and market segmentation. In many empirical settings, the analyst cannot perfectly explain taste heterogeneity in terms of observed individual characteristics, and taste heterogeneity remains to a substantial extent random from the analyst's point-of-view \citep[e.g.][]{bhat_accommodating_1998}. If decision-makers are assumed to employ a decision strategy that is consistent with random utility maximisation, a mixed random utility model such as the mixed multinomial logit (M-MNL) model or the mixed multinomial probit (M-MNP) model can accommodate any empirical random heterogeneity distribution by marginalising the discrete choice kernel over some mixing distribution, which describes the unobserved distribution of tastes in the sample \citep{mcfadden_mixed_2000,train_discrete_2009}. However, the ability of mixed random utility models to recover any true heterogeneity distribution is only predicated on an existence proof \citep[see][]{mcfadden_mixed_2000} and therefore, the analyst is required to select an appropriate mixing distribution to capture an unobserved heterogeneity distribution in a given empirical setting. 

There are two principal approaches to account for unobserved taste heterogeneity in mixed random utility models \citep[e.g.][]{wedel_discrete_1999}: Parametric mixed random utility models are based on the assumption that individual taste parameters are drawn from a sample-level continuum of tastes with a specific shape. In nonparametric mixed random utility models, individuals are probabilistically assigned to a countable, typically finite number of segments with homogeneous tastes. If the stochastic error terms of the utilities are assumed to be independent and identically distributed Gumbel random variates, the former approach can be referred to as parametric mixed multinomial logit (PM-MNL) model \citep[e.g.][]{mcfadden_mixed_2000,train_discrete_2009}, while a standard implementation of the latter approach is known as latent class multinomial logit (LC-MNL) model \citep[e.g.][]{bhat_endogenous_1997,greene_latent_2003,kamakura_probabilistic_1989}.\footnote{The remainder of this contribution focusses on heterogeneity representations within the M-MNL model, but for completeness, we point out that flexible representations of unobserved taste heterogeneity can also be accommodated by the M-MNP model \citep[see][]{bhat_new_2017,bhat_new_2012}.}

Both the PM-MNL and LC-MNL models are widely used in disciplines studying individual choice behaviour but are subject to limitations: The distributional assumptions of parametric mixture models may be overly rigid and may not yield convincing representations of inter-individual random taste heterogeneity, as the shape of the estimated taste parameter distribution is constrained to be equal to the functional form of the imposed parametric random distribution \citep{vij_random_2017}. For valid inferences in PM-MNL models, it is therefore imperative to correctly specify the mixing distribution of the randomised taste parameters \citep{hess_estimation_2005}. In practice however, the analyst is unlikely to be able to exhaust the hypothesis space of theoretically-feasible parametric distribution functions \citep[e.g.][]{keane_comparing_2013}. Nonparametric mixture approaches such as the LC-MNL model free the analyst from rigid distributional assumptions \citep[e.g.][]{greene_latent_2003} but are cumbersome to estimate, as the number of mixture components needs to be determined exogenously. 

\subsection{Existing work}

In response to the limitations of the PM-MNL and the LC-MNL models, further semi-nonparametric and nonparametric variations of the M-MNL models have been proposed \citep[see][for reviews]{vij_random_2017,yuan_guide_2015}. 

Several semi-nonparametric approaches combine discrete and continuous representations of unobserved heterogeneity by employing mixing distributions that are finite mixtures of continuous distributions. For example, \citet{bujosa_combining_2010}, \citet{fosgerau_comparison_2009} and \citet{keane_comparing_2013} use mixing distributions that are finite mixtures of independent normal distributions. Similarly, \citet{greene_revealing_2013} employ a finite mixture of independent triangular distributions as a random taste parameter distribution. \citet{train_em_2008} allows for correlation between random taste parameters within mixture components by leveraging a finite mixture of multivariate Gaussians as a mixing distribution. Semi-nonparametric approaches relying on mixing distributions that are finite mixtures of continuous distributions are conceptually appealing, as unbounded continuous distributions can be closely approximated by a finite mixture of multivariate Gaussians \citep[e.g.][]{keane_comparing_2013}. However, such semi-nonparametric approaches are computationally demanding, as the analyst must perform post-hoc model selection to determine the appropriate number of mixture components.

Another class of semi-nonparametric approaches leverages flexible functionals such as Legendre polynomials \citep[see][]{fosgerau_practical_2007} and B-splines \citep[see][]{bastin_estimating_2010} to accommodate unobserved taste heterogeneity in M-MNL models. These approaches allow for flexible representations of unobserved heterogeneity but require the analyst to configure the complexity of the employed functional prior to estimation. Relatedly, \citet{train_mixed_2016} proposes a semi-nonparametric M-MNL model where an additional discrete mixing distribution is imposed on the parameters of a variety of flexible functionals such as step, spline or polynomial functions. The framework can flexibly recover differently-shaped taste parameter distributions but requires the analyst to select both the complexity of the discrete mixing distribution for the parameters of the functional as well as the functional itself prior to estimation. 

The LC-MNL model is the simplest implementation of a M-MNL model with a nonparametric discrete mixing distribution with a finite number of support points, whose locations and associated probability mass need to be estimated. In practice, high-dimensional LC-MNL models are often plagued by identification issues due to the multi-modality of the log-likelihood function. Therefore, a stream of literature proposes nonparametric mixing distributions with structured support points \citep[see][]{dong_comparison_2014,train_em_2008,vij_random_2017}. Effectively, these approaches implement multi-variate histogram estimators by defining a multi-dimensional grid on the coefficients space such that the heterogeneity distribution in question can be closely approximated by estimating the amount of probability mass positioned on the vertices of the grid. M-MNL models with gridded mixing distributions can capture complex heterogeneity distributions \citep[see e.g.][]{vij_random_2017} but require the analyst to specify the complexity of the grid of mass points a priori. 

Extant M-MNL models with nonparametric discrete mixing distributions including the LC-MNL model are finite-dimensional nonparametric models, which rely on an a priori specification of a finite, comparatively large parameter space to flexibly cover the desired hypothesis space of heterogeneity distributions. \citet{train_em_2008} therefore suggests to refer to finite-dimensional nonparametric models as ''super-parametric`` models due to their large number of parameters. However, to be precise, finite-dimensional and infinite-dimensional nonparametric models can be distinguished \citep{gelman_bayesian_2013}. The defining characteristic of infinite-dimensional nonparametric models is that the complexity of a model, i.e. the size of the parameter space, is endogenised \citep{gelman_bayesian_2013}. Such infinite-dimensional models are known as Bayesian nonparametric models, because model complexity is incorporated into the posterior density via stochastic process priors and estimated conditional on the observed data \citep{gershman_tutorial_2012,orbanz_bayesian_2011}. 

\subsection{Approach and objective}

Dirichlet process mixture models \citep{antoniak_mixtures_1974} are a flexible class of Bayesian nonparametric models, which preclude the a priori specification of the number mixture components in a discrete mixture model by exploiting the properties of the Dirichlet process \citep{ferguson_bayesian_1973}. The Dirichlet process is a stochastic process, whose realisations are probability distributions. The Dirichlet process exhibits two useful properties: First, realisations from a Dirichlet process are discrete and second, repeated samples from a realisation from a Dirichlet process are clustered with non-zero probability, while the expected number of clusters grows only logarithmically in the sample size \citep[e.g.][]{gershman_tutorial_2012,teh_dirichlet_2011}. When used as a nonparametric prior in discrete mixture models, the Dirichlet process induces a random partition where each individual cluster of the partition is characterised by its own parameter vector for the probability distribution of the response variable. In such Dirichlet process mixture models, the number of mixture components need not be fixed a priori, as it is inferred from the evidence. Because the complexity of the discrete mixing distribution is not fixed prior to estimation, Dirichlet process mixture models can be conceived as infinite mixture models \citep{rasmussen_infinite_1999}. The truncated stick-breaking construction \citep{ishwaran_gibbs_2001} is a finite-dimensional, precise approximation of the Dirichlet process and permits tractable inference in Dirichlet process mixture models.

In this paper, we present a M-MNL model, which leverages the truncated stick-breaking process representation of the Dirichlet process as a flexible nonparametric mixing distribution. The proposed model is a Dirichlet process mixture multinomial logit (DPM-MNL) model and accommodates discrete representations of heterogeneity, like a LC-MNL model. However, unlike a LC-MNL model, the proposed DPM-MNL model does not require the analyst to fix the number of mixture components prior to estimation, as the complexity of the discrete mixing distribution is inferred from the evidence. For posterior inference in the proposed DPM-MNL model, we derive an expectation maximisation (EM) algorithm \citep{dempster_maximum_1977}. In a simulation study, we demonstrate that the proposed DPM-MNL model can flexibly capture differently-shaped taste parameter distributions. Furthermore, we empirically validate the model framework in a case study on motorists' route choice preferences and find that the proposed model outperforms an LC-MNL model and PM-MNL models with common continuous mixing distributions in terms of in-sample fit and out-of-sample predictive ability. Compared to extant modelling approaches, the proposed DPM-MNL model substantially abbreviates specification searches, as it relies on less restrictive parametric assumptions and does not require the analyst to specify the complexity of its discrete mixing distribution a priori.

\subsection{Contribution}

Dirichlet process mixture models with kernels for continuous dependent data have been presented in the domains of statistics and machine learning \citep[see the literature referenced in][]{teh_dirichlet_2011}. In the context of discrete choice analysis, Dirichlet process mixture modelling methods have been leveraged to construct discrete choice models with discrete \citep{ansari_semiparametric_2006,kim_assessing_2004} and continuous \citep{burda_bayesian_2008,li_bayesian_2013} representations of taste heterogeneity. \citet{kim_assessing_2004} present Dirichlet process mixture models with multinomial logit and multinomial probit kernels. Similarly, \citet{ansari_semiparametric_2006} allow for a discrete representation of heterogeneity by defining taste parameters as a Dirichlet process mixture in a variety of Thurstonian models including the multinomial probit model. \citet{burda_bayesian_2008} accommodate continuous representations of taste heterogeneity in a Dirichlet process mixture models, where the kernels are mixed logit models with multivariate normal heterogeneity distributions. \citet{li_bayesian_2013} present a mixed multinomial probit model, where the mixing distribution of some coefficients is a Dirichlet process mixture of normal distributions. 

All of the four applications of Dirichlet process mixture modelling methods in discrete choice analysis employ Markov Chain Monte Carlo methods for model inference and present case studies in the realms of consumer behaviour. Relative to this literature, the DPM-MNL model presented in the current paper affords two innovations: First, we demonstrate that an EM algorithm can be used for posterior inference in a Dirichlet process mixture model of discrete choice. Second, we empirically validate the proposed model framework in a case study on travel demand and comprehensively benchmark the proposed DPM-MNL model against established M-MNL models with both continuous and discrete mixing distributions in terms of both in-sample fit and out-of-sample predictive ability. 

Moreover, our paper contributes to two other strands of literature: First, we introduce a Dirichlet process mixture model with a multinomial logit kernel into the domain of behavioural travel demand analysis and demonstrate the value of the proposed model framework in a case study on motorists' route choice preferences. Second, we contribute to a growing body of literature concerned with the development and application of EM algorithms for the estimation of complex discrete choice models \citep{bhat_endogenous_1997,sohn_expectation-maximization_2016,train_em_2008,train_discrete_2009,vij_random_2017}. While it is well known that the EM algorithm can facilitate inference in finite-dimensional discrete mixture M-MNL models \citep{bhat_endogenous_1997,train_em_2008,train_discrete_2009,vij_random_2017}, we demonstrate that the computational benefits of the EM algorithm generalise to the infinite-dimensional setting.

\subsection{Paper outline}

The remainder of this paper assumes the following structure: The subsequent section provides conceptual and technical prerequisites to the proposed Dirichlet process mixture model of discrete choice. The formulation of the proposed model framework is presented in Section \ref{S_modelFramework}, and Section \ref{S_modelInf} explicates the inference approach. The simulation and case studies are presented in Sections \ref{S_simStudy} and \ref{S_caseStudy}. Section \ref{S_conclusion} concludes by summarising the proposed modelling approach, by acknowledging limitations and by pointing at directions for future research.

\section{Prerequisites} \label{S_background}

In this section, we provide the conceptual and technical prerequisites to the proposed Dirichlet process mixture model of discrete choice. First, we motivate Dirichlet process mixture models by the desideratum to obviate the a priori specification of the number of mixture components in a finite mixture model (Section \ref{S_desideratum}). Next, we introduce the Dirichlet process, which is foundational to Dirichlet process mixture models (Section \ref{S_dp}), and describe alternative representations of the Dirichlet process (Section \ref{S_dp_alt}). These alternative representations illustrate the clustering and discreteness properties of the Dirichlet process and are necessary for applications of the Dirichlet process in statistical models. Ultimately, we present the generative process of an exemplative Dirichlet process mixture model (Section \ref{S_dpm}).  

\subsection{Desideratum: An infinite mixture model} \label{S_desideratum}

In finite mixture models, analytical units are probabilistically assigned to one and only one of a finite number of mixture components, each of which is characterised by its own parameter vector for the component-specific probability distribution of the dependent variable. For example, in the LC-MNL model, decision-makers are distributed over a finite number of taste segments, each of which has its own taste vector parameterising the component-specific multinomial logit (MNL) kernel, from which the observed choices are drawn. The LC-MNL model assumes the following data generating process (DGP): 
\begin{align}
q_{n} & \sim \mbox{Categorical}(\boldsymbol{\pi}) & & n = 1,\dots,N \label{eq_lcMnl1} \\
y_{n,t} & \sim \mbox{MNL} (\boldsymbol{\beta}_{k}, \boldsymbol{X}_{n,t}) & & n = 1,\dots,N,  \quad t = 1,\dots,T_{n}, \label{eq_lcMnl2}
\end{align}
where $n \in \{1,\dots,N\}$ indexes individuals and $t \in \{1,\dots,T_{n}\}$ indexes choice occasions for individual $n$. $q_{n} \in \{1,\ldots,K\}$ denotes an individual's component assignment, which is drawn from a categorical distribution with parameter $\boldsymbol{\pi}$. The random variable $y_{n,t}$ denotes the chosen alternative and takes values in the set of available alternatives $C_{n,t}$. $y_{n,t}$ is assumed to be drawn from an MNL kernel. Thence, the probability that individual $n$ chooses alternative $j \in C_{n,t}$ on choice occasion $t$ is 
\begin{equation}
P(y_{n,t} = j \vert \boldsymbol{\beta}_{k}, \boldsymbol{X}_{n,t}) = \frac{\exp( V(\boldsymbol{X}_{n,t,j}, \boldsymbol{\beta}_{k}))}{\sum_{j \in C_{n,t}}\exp(V(\boldsymbol{X}_{n,t,j'}, \boldsymbol{\beta}_{k}))},
\end{equation}
where $\boldsymbol{\beta}_{k}$ is the component-specific taste vector. $\boldsymbol{X}_{n,t}$ is a matrix of covariates and $\boldsymbol{X}_{n,t,j}$ is a row of $\boldsymbol{X}_{n,t}$. $V(\boldsymbol{X}_{n,t,j}, \boldsymbol{\beta}_{k})$ is a function giving the deterministic component of utility.

In finite mixture models, the number of mixture components $K$ represents an exogenous model parameter. For this reason, the final specification of a finite mixture model must be determined based on a consideration of post-hoc model selection criteria, which are applied to a candidate set of models with varying numbers of mixture components. As such specification searches often require considerable time and effort, we would like to obviate the a priori specification of the number of mixture components in a discrete mixture model without compromising on the distributional flexibility afforded by a nonparametric mixing distribution. A discrete mixture model where the number of mixture components need not be fixed a priori, as it is inferred from the evidence, is an infinite mixture model \citep{rasmussen_infinite_1999}. Such a model can be realised with the help of the Dirichlet process \citep{ferguson_bayesian_1973}. 

\subsection{Dirichlet process} \label{S_dp}

To allow a mixture model to endogenously adapt the complexity of its discrete mixing distribution to the evidence, we can leverage the properties of the Dirichlet process \citep{ferguson_bayesian_1973}, which is a stochastic process, whose draws are probability distribution over some measurable space $\Theta$ \citep[see][for extended discussions of the Dirichlet process]{gelman_bayesian_2013,gershman_tutorial_2012,teh_dirichlet_2011}. A Dirichlet process is parameterised by a positive, real-valued concentration parameter $\alpha$ and a base measure $\mbox{G}_{0}$, which itself is a probability distribution on $\Theta$. In that vein, we let $\mbox{G} \sim \mbox{DP}(\alpha,\mbox{G}_{0})$ denote a sample from a Dirichlet process. 

The Dirichlet process is an infinite-dimensional generalisation of the Dirichlet distribution (which gives a probability simplex associated with the outcomes of a multinomial event); \citet{ferguson_bayesian_1973} shows that if $\mbox{G} \sim \mbox{DP}(\alpha,\mbox{G}_{0})$, any finite, mutually exclusive partition $A_{1}, \ldots, A_{k}$ of a space $\Theta$ exhibits a Dirichlet distribution, i.e.
\begin{equation} \label{eq_dp_formal}
\mbox{G}(A_{1}), \ldots, \mbox{G}(A_{k}) \sim \mbox{Dirichlet} \left ( \alpha \mbox{G}_{0}(A_{1}), \ldots, \alpha \mbox{G}_{0}(A_{k}) \right ).
\end{equation}
$\Theta$ may be any measurable space such as the $n$-dimensional real space and $\mbox{G}$ defines how the probability mass is distributed over the partitioned space, i.e. $\mbox{G}(A_{l})$ is the amount of probability mass in region $A_{l}$. $\mbox{G}_{0}$ is an initial guess about $\mbox{G}$ and $\alpha$ controls the proximity of $\mbox{G}_{0}$ and $\mbox{G}$. 

\begin{figure}[H]
\centering
\includegraphics[width = 0.8 \textwidth]{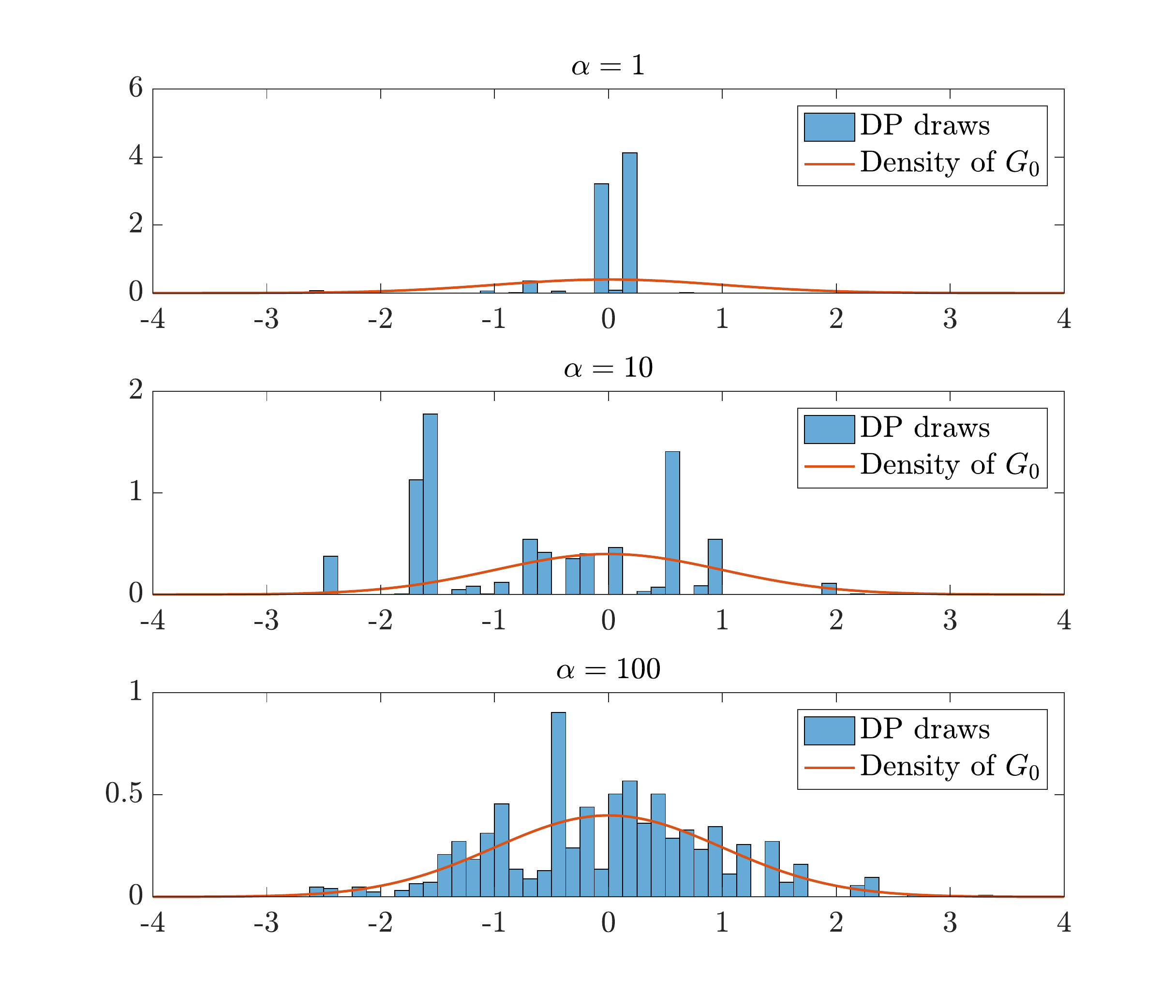}
\caption{Distribution of 1,000 draws from the Dirichlet process based on the stick-breaking process construction of the Dirichlet process with baseline $\mbox{N}(0,1)$ for different values of $\alpha$} \label{f_dp}
\end{figure}

The mean of a Dirichlet process is its baseline distribution \citep{teh_dirichlet_2011}, i.e. $\mathbf{E} \left [ \mbox{G}(A) \right ]  = \mbox{G}_{0}(A)$. The concentration parameter $\alpha$ controls the variance of the Dirichlet process \citep{teh_dirichlet_2011}, i.e. $\mbox{Var} \left [ \mbox{G}_{0}(A) \right ]  = \frac{\mbox{G}_{0}(A) \left( 1 - \mbox{G}_{0}(A) \right )}{\alpha + 1}$. For $\alpha \rightarrow 0$, draws from the Dirichlet process agglomerate around the mean. For $\alpha \rightarrow \infty$, $G \rightarrow \mbox{G}_{0}$, i.e. the Dirichlet process approaches the baseline distribution. Figure \ref{f_dp} illustrates the behaviour of a Dirichlet process with a standard normal base measure for different values of $\alpha$.\footnote{Note that the draws in Figure \ref{f_dp} were generated with the help of the stick-breaking process construction \citep{ishwaran_gibbs_2001,sethuraman_constructive_1994} of the Dirichlet process. We will introduce this alternative representation of the Dirichlet process in Section \ref{S_stick}, once the theory has been further developed.} 

The Dirichlet process exhibits two important properties, which qualify it as a nonparametric prior in clustering and segmentation models: First, realisations from the Dirichlet process are discrete and second, repeated samples from a realisation $\mbox{G}$ from the Dirichlet process are clustered with non-zero probability. \citet{ferguson_bayesian_1973} provides formal proofs for both properties. However, the properties of the Dirichlet process can also be illustrated in more intuitive ways through alternative representations of the Dirichlet process.

\FloatBarrier
\subsection{Alternative representations of the Dirichlet process} \label{S_dp_alt}

The formal definition of the Dirichlet process (\ref{eq_dp_formal}) is not immediately useful for practical applications. For this reason, three alternative representations of the Dirichlet process have been proposed, namely the Blackwell-MacQueen urn scheme \citep{blackwell_ferguson_1973}, the Chinese Restaurant process \citep{aldous_exchangeability_1985} and the stick-breaking process construction \citep{sethuraman_constructive_1994}. The Blackwell-MacQueen urn scheme and the Chinese Restaurant process are closely related to one another and are helpful in building intuition about the clustering property of the Dirichlet process. The stick-breaking construction illustrates discreteness property of the Dirichlet process and is leveraged in the proposed Dirichlet process mixture model of discrete choice.

\FloatBarrier
\subsubsection{Blackwell-MacQueen urn scheme and Chinese Restaurant process} \label{S_crp}

The Blackwell-MacQueen urn scheme \citep{blackwell_ferguson_1973} elucidates the clustering property of the Dirichlet process. Consider the following DGP:
\begin{align}
\mbox{G} & \sim \mbox{DP}(\alpha,\mbox{G}_{0}) \\
 \theta_{n} & \sim \mbox{G} & & n = 1,\dots,N,
\end{align}
i.e. repeated draws $\theta_{1:N}$ are taken from a realisation $\mbox{G}$ from a Dirichlet process. \citet{blackwell_ferguson_1973} show that $\theta_{1:N}$ constitute a P\`olya sequence: Conditional on previous draws $\theta_{1:n-1}$, the probability that a new draw $\theta_{n} \sim \mbox{G}$ assumes a new value $\theta^{*} \sim \mbox{G}_{0}$ is $P(\theta_{n} = \theta^{*} \vert \theta_{1:n-1}) = \frac{\alpha}{\alpha + n - 1}$, and the probability that $\theta_{n}$ assumes an existing value $\theta_{k}$ is $P(\theta_{n} = \theta_{k} \vert \theta_{1:n-1}) = \frac{n_{k}}{\alpha + n - 1}$, where $n_{k}$ denotes the number of times $\theta_{k}$ appears in $\theta_{1:n-1}$. The latter probability is non-zero so that $\theta_{1:N}$ are clustered with non-zero probability. We also observe that the probability that $\theta_{n}$ assumes an existing value $\theta_{k}$ is proportional to $n_{k}$. Hence, clustering under the Dirichlet process is subject to preferential attachment and clusters that are large are relatively more likely to grow in size, as new draws are taken. On the other hand, the probability that $\theta_{n}$ assumes a new value is proportional to the concentration parameter $\alpha$: So the larger $\alpha$ is, the more likely are draws distributed over different clusters (see Figure \ref{f_dp}). Furthermore, it can be shown that the expected number of distinct clusters $\mathbf{E}(K)$ under a Dirichlet process only grows logarithmically in the sample size $N$ and is proportional to $\alpha$, i.e. $\mathbf{E}(K) = \alpha \ln N$ for $\alpha < \frac{N}{\ln N}$ \citep[see e.g.][]{gershman_tutorial_2012,teh_dirichlet_2011}. 

\begin{figure}[h]
\centering
\includegraphics[width = 0.9 \textwidth]{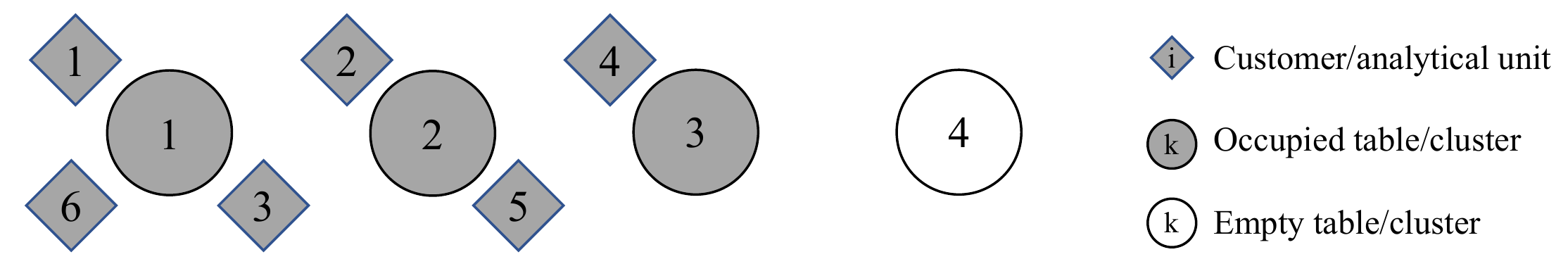}
\caption{Illustration of a partition induced by the Chinese Restaurant process} \label{figure_crp}
\end{figure}

Ergo, the Dirichlet process induces a random partition where the individual clusters of the partition are each characterised by a specific realisation $\theta_{k}$ from $\mbox{G}_{0}$. The Chinese Restaurant process representation of the Dirichlet process \citep{aldous_exchangeability_1985} induces a random partition like the Blackwell-MacQueen urn scheme, but does not assign draws $\theta_{k} \sim \mbox{G}_{0}$ to the clusters. The Chinese Restaurant process representation receives its name from a metaphor describing the process of sequentially seating customers in a restaurant and provides a more vivid illustration of the clustering property of the Dirichlet process: Consider a restaurant with an infinite number of tables, each of which provides seating for an infinite number of customers. The first customer entering the restaurant sits at any table. The second customer takes a seat at the same table as the first customer with probability $\frac{1}{1 + \alpha}$ and at another table with probability $\frac{\alpha}{1 + \alpha}$. The $n$th customer sits at an occupied table with probability proportional to the number of customers already seated at the table and at an unoccupied table with probability proportional to $\alpha$. More formally, the Chinese Restaurant process can be represented as follows: The probability of assignment $q_{n}$ of analytical unit $n \in \{1,\ldots,N \}$ to cluster $k \in \{1,\ldots,K \}$ is $P(q_{n} = k \vert q_{1:n-1}) = \frac{n_{k}}{\alpha + n - 1}$, if $n$ joins existing cluster $k$ with size $n_{k}$, and $P(q_{n} = k \vert q_{1:n-1}) = \frac{\alpha}{\alpha + n - 1}$, if $n$ initiates a new cluster. Figure \ref{figure_crp} illustrates a partition induced by the Chinese Restaurant process.

\FloatBarrier
\subsubsection{Stick-breaking process} \label{S_stick} 

\citet{ferguson_bayesian_1973} shows that the realisations from the Dirichlet process are probability-weighted point masses, i.e.
\begin{equation} \label{eq_atomic}
\mbox{G} = \sum_{k=1}^{\infty} \pi_{k} \delta_{\theta_{k}},
\end{equation}
where 
$\pi_{k} \in [0,1]$ with $\sum_{k=1}^{\infty} \pi_{k} = 1$ is a probability weight and $\delta_{\theta_{k}}$ is a probability point mass centred at $\theta_{k} \sim \mbox{G}_{0}$. This atomic representation of a realisation from the Dirichlet process demonstrates the discreteness property of the Dirichlet process and is exploited by the stick-breaking process construction \citep{sethuraman_constructive_1994} of the Dirichlet process.

\begin{figure}[h]
\centering
\includegraphics[width = 0.9 \textwidth]{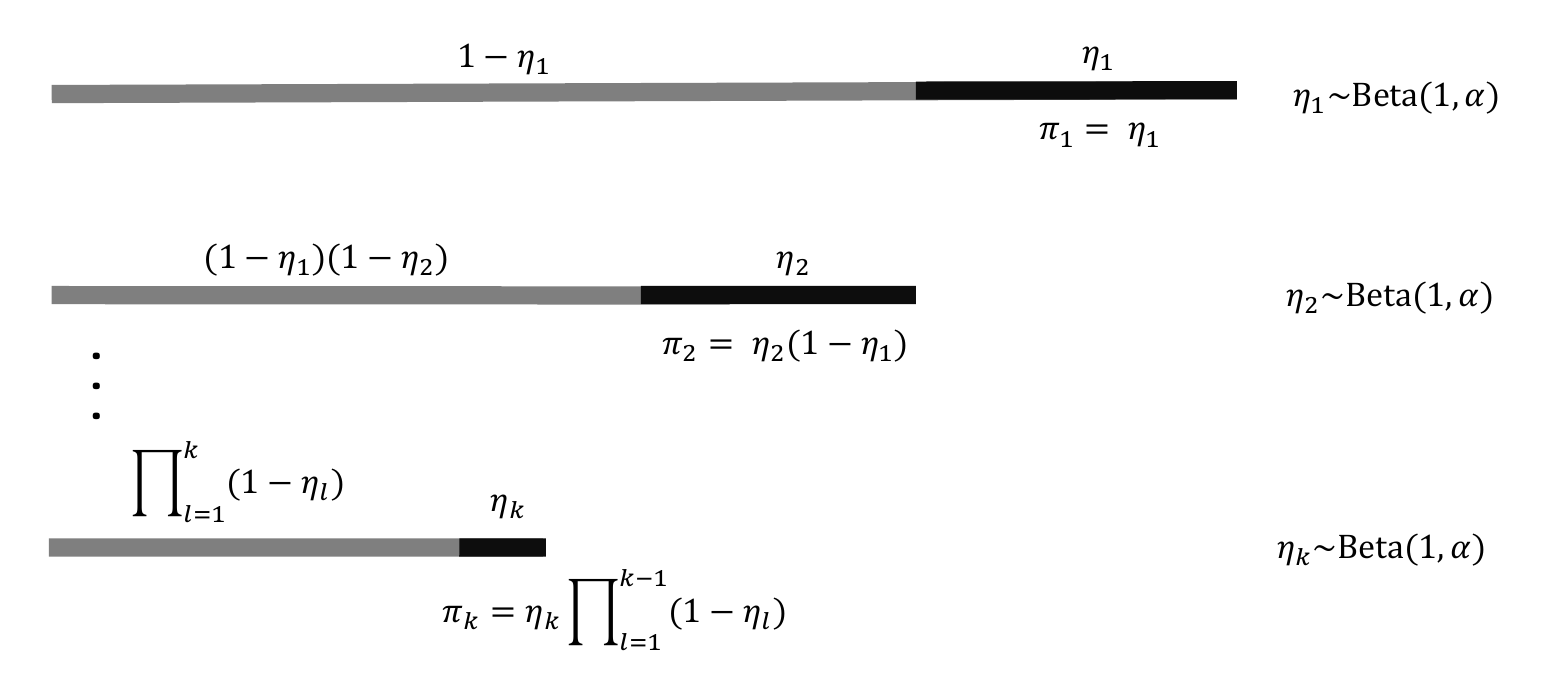}
\caption{Illustration of the stick-breaking process construction of the Dirichlet process} \label{figure_stick-breaking}
\end{figure}

The stick-breaking process construction of the Dirichlet process defines a realisation from a Dirichlet process $\mbox{G} \sim \mbox{DP}(\alpha, \mbox{G}_{0})$ as a discrete mixture of point masses, whereby the component weights are factorisations of Beta-distributed random variables, i.e.
\begin{equation}
\mbox{G} = \sum_{k=1}^{\infty} \pi_{k} \delta_{\theta_{k}}
\end{equation}
with 
\begin{equation} \label{e_stick-breaking}
\theta_{k} \sim \mbox{G}_{0},
\quad
\eta_{k} \sim \text{Beta}(1,\alpha),
\quad
\pi_{k} = \eta_{k} \prod_{l=1}^{k-1}(1 - \eta_{l}),
\quad k = 1,\ldots,\infty,
\end{equation}
where 
$\pi_{k} \in [0,1]$  with $\sum_{k=1}^{\infty} \pi_{k} = 1$ is a probability weight and $\delta_{\theta_{k}}$ is the associated point mass centred at $\theta_{k}$, which is a unique realisation from $\mbox{G}_{0}$. The stick-breaking process receives its name from the metaphor describing the process of breaking a stick of unit length into an infinite number of pieces \citep[e.g.][]{gelman_bayesian_2013,teh_dirichlet_2011}. Figure \ref{figure_stick-breaking} illustrates the stick-breaking process: Beginning with a stick of unit length, we break the stick at $\eta_{1} \sim \text{Beta}(1,\alpha)$ and assign $\pi_{1}$ to the piece we broke off. We draw $\eta_{2} \sim \text{Beta}(1,\alpha)$ and break a piece of size $\pi_{2} = \eta_{2} (1 - \eta_{1})$ from the remaining $1 - \eta_{1}$ stick. Subsequently, we continue to break off sticks of sizes $\pi_{3}, \ldots , \pi_{k}$  of the remainder of the stick. 

The truncated stick-breaking process \citep{ishwaran_gibbs_2001} is a finite-dimensional approximation of the infinite-dimensional stick-breaking process. Under the truncated stick-breaking process representation, $\mbox{G}$ is given by
\begin{equation} 
\mbox{G} \approx \sum_{k=1}^{K} \pi_{k} \delta_{\theta_{k}},
\end{equation}
with 
\begin{align} \label{e_stick-breaking}
\theta_{k} & \sim \mbox{G}_{0} && \quad k = 1,\ldots,K \\
\eta_{k} & \sim \text{Beta}(1,\alpha), \quad \pi_{k} = \eta_{k} \prod_{l=1}^{k-1}(1 - \eta_{l}) && \quad k = 1,\ldots,K - 1
\end{align}
where the truncation level $K$ is chosen by the analyst. To assure that $\sum_{k=1}^{K} \pi_{k} = 1$, the final random variable is degenerate, i.e. $\eta_{K} = 1$ so that $\pi_{K} = 1 - \sum_{k=1}^{K - 1} \pi_{k}$. This stick-breaking construction of a probability vector can be referred to as Griffiths-Engen-McCloskey (GEM) distribution \citep{pitman_sequential_2006}. We write $\boldsymbol{\pi} \sim \mbox{GEM}_{K}(\alpha)$ to denote that the probability vector $\boldsymbol{\pi}$ is realisation from a truncated stick-breaking process with concentration parameter $\alpha$ and truncation level $K$.

At first glance, the use of a truncated stick-breaking process prior appears to defeat the purpose of the Bayesian nonparametric modelling paradigm, as we are essentially defining a finite mixture model of dimension $K$. However, the truncated stick-breaking process prior induces a shrinkage on the number of effectively populated mixture components, while maintaining the computational advantages of a finite mixture model \citep{gelman_bayesian_2013}. In fact, the residual probability $\pi_{K}$ is negligibly small for reasonably large $K$ and most $\alpha$ values that are encountered in practice \citep{ishwaran_markov_2000,ohlssen_flexible_2007}. In Section \ref{S_practicalities}, we provide a detailed discussion about the choice of $K$.

For inference in the proposed Dirichlet process mixture model of discrete choice, we exploit the fact that the truncated stick-breaking representation of the Dirichlet process is a generalised Dirichlet distribution \citep{connor_concepts_1969}, which is the joint distribution of $K-1$ independent Beta-distributed random variables: Let
\begin{equation}
\eta_{k} \sim \text{Beta}(a_{k},b_{k}),
\quad k = 1,\ldots,K - 1,
\end{equation}
and $\eta_{K} = 1$. Furthermore, define $\pi_{k} = \eta_{k} \prod_{l=1}^{k-1}(1 - \eta_{l})$. Then, the joint density of $\pi_{1:K}$ is given by 
\begin{equation}
f(\pi_{1:K} \vert a_{1:(k-1)},b_{1:(k-1)}) =
\left (
\prod_{k=1}^{K-1} B(a_{k},b_{k})
\right )^{-1}
\prod_{k=1}^{K-1} \pi_{k}^{a_{i} - 1} \left ( \sum_{k'=k}^{K-1} \pi_{k'}^{b_{i-1} - (a_{i} + b_{i})} \right )
\pi_{K}^{b_{K-1} - 1},
\end{equation}
$B(a,b)$ denotes the Beta function evaluated at $\{a,b\}$. The joint density  $\pi_{1:K}$ under the truncated stick-breaking process is obtained by letting $a_{k} = 1$ and $b_{k} = \alpha$ $\forall$ $k = 1,\ldots,K - 1$. The generalised Dirichlet distribution is the conjugate prior of the multinomial distribution. The compound distribution of this conjugate pair is known as generalised-Dirichlet-multinomial distribution and assumes the following DGP: 
\begin{align}
\boldsymbol{\pi} & \sim \mbox{Generalised-Dirichlet}(\pi_{1:K} \vert a_{1:(k-1)},b_{1:(k-1)}) \\
\boldsymbol{x} & \sim \mbox{Multinomial}(\boldsymbol{x} \vert \boldsymbol{\pi}),
\end{align}
where $\boldsymbol{x}$ is a $K$-dimensional vector of category counts. The marginal density of $\boldsymbol{x}$ is given by \citep[e.g.][]{zhou_mm_2010}:
\begin{equation}
\begin{split}
 f(\boldsymbol{x} \vert a_{1:(k-1)},b_{1:(k-1)}) & = \int f(\boldsymbol{x} \vert \boldsymbol{\pi}) f(\boldsymbol{\pi} \vert a_{1:(k-1)},b_{1:(k-1)}) d \boldsymbol{\pi} \\
& = \frac{\Gamma ( N + 1 )}{\prod_{k=1}^{K} \Gamma ( x_{k} + 1 )} \prod_{k=1}^{K-1} \frac{\Gamma (a_{k} + x_{k}) \Gamma (b_{k} + m_{k+1}) \Gamma (a_{k} + b_{k})}{\Gamma (a_{k}) \Gamma (b_{k}) \Gamma (a_{k} + b_{k} + m_{k} )}
\end{split}
\end{equation}
where $N = \sum_{k=1}^{K} x_{k}$ and $m_{k} = \sum_{k'=k}^{K} x_{k'}$. $\Gamma(\cdot)$ denotes the Gamma function.

\subsection{Dirichlet process mixture models} \label{S_dpm}

The Dirichlet process can be used as a nonparametric prior in mixture models to construct Dirichlet process mixture models \citep{antoniak_mixtures_1974}. In Dirichlet process mixture models, the number of mixture components is not fixed, but is endogenously determined based on the evidence \citep[see][for general discussions of Dirichlet process mixture models]{blei_variational_2006,gelman_bayesian_2013,mcauliffe_nonparametric_2006}. Staying with the example of a mixture model with MNL kernels (\ref{eq_lcMnl1}--\ref{eq_lcMnl2}), we can write out the generative process of an exemplative Dirichlet process mixture model:
\begin{align}
 \mbox{G} & \sim \mbox{DP}(\alpha,\mbox{G}_{0}) \label{eq_genDpm1} \\
\boldsymbol{\beta}_{n} & \sim \mbox{G} & & n = 1,\dots,N \\
y_{n,t} & \sim \mbox{MNL} (\boldsymbol{\beta}_{n}, \boldsymbol{X}_{n,t}) & & n = 1,\dots,N,  \quad t = 1,\dots,T_{n}, \label{eq_genDpm2} 
\end{align}
where
\begin{equation}
P(y_{n,t} = j \vert \boldsymbol{\beta}_{n}, \boldsymbol{X}_{n,t}) = \frac{\exp(V (\boldsymbol{X}_{n,t}, \boldsymbol{\beta}_{n}))}{\sum_{j \in C_{n,t}}\exp(V (\boldsymbol{X}_{n,t}, \boldsymbol{\beta}_{n}))}.
\end{equation}
In practical terms, the Dirichlet process prior in the generative process (\ref{eq_genDpm1}--\ref{eq_genDpm2}) defines an infinite mixture model by precluding the a priori specification of the number of mixture components. Due to the clustering property of the Dirichlet process, $\boldsymbol{\beta}_{1:N}$ are clustered with non-zero probability and the sample can be partitioned ex post into countable segments based on the distinct component-specific parameter values \citep[e.g.][]{blei_variational_2006}. 

\section{Model framework} \label{S_modelFramework}

We now present the formulation of a Dirichlet process mixture model, where the component-specific probability distribution functions are MNL kernels.\footnote{For completeness, we point out that it is straightforward to generalise our proposed model framework including the corresponding inference approach presented in Section \ref{S_modelInf} to other kernels that are commonly considered in discrete choice analysis.} Our model formulation involves the truncated stick-breaking process construction of the Dirichlet process as a computationally efficient means to allow the proposed discrete choice model to adapt the complexity of its discrete mixing distribution to the evidence.

\subsection{Generative process}

\begin{figure}[h]
	\centering
	\subcaptionbox{DPM-MNL model \label{f_dpm_mnl}} 
  	[.48\linewidth]{\resizebox{.4\textwidth}{!}{\input{model_infinite-mm}}}
\subcaptionbox{LC-MNL model \label{f_lc_mnl}} 
  	[.48\linewidth]{\resizebox{.4\textwidth}{!}{\input{model_finite-mm}}}
\caption{Generative processes of DPM-MNL and LC-MNL models}
\end{figure}
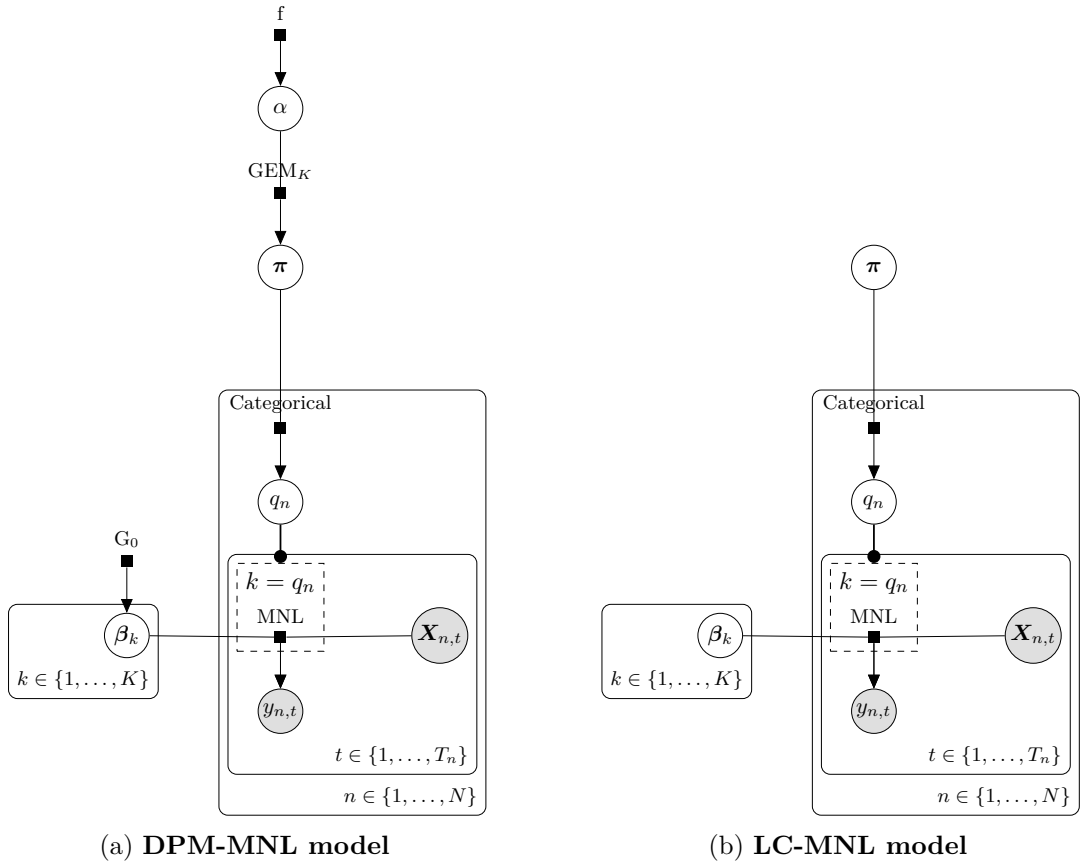

The generative process of the proposed Dirichlet process mixture multinomial logit (DPM-MNL) model is visualised in Figure \ref{f_dpm_mnl} and can be described as follows: Decision-makers are indexed by $n \in \{1, \ldots, N \}$ and are distributed over $K$ mixture components indexed by $k \in \{1, \ldots, K \}$. Each mixture component is characterised by a taste vector $\boldsymbol{\beta}_{k}$ with prior $\mbox{G}_{0}$.  The latent variable $q_{n}$ indicates a decision-maker's component allocation such that $q_{n} = k$, if decision-maker $n$ is assigned to mixture component $k$. $q_{n}$ controls an individual's tastes $\boldsymbol{\beta}_{k}$ such that an observed choice $\boldsymbol{y}_{n,t}$ is a function of tastes $\boldsymbol{\beta}_{k}$ and covariates $\boldsymbol{X}_{n,t}$. $q_{n}$ is a realisation from a categorical distribution with parameter $\boldsymbol{\pi}$, which in turn is obtained via a $\mbox{GEM}_{K}$ distribution with truncation level $K$ and concentration parameter $\alpha$. $\alpha$ is drawn from some distribution f.

Stated succinctly, the generative process of the DPM-MNL model is:
\begin{align}
\alpha & \sim \mbox{f}, \\
\boldsymbol{\pi} & \sim \mbox{GEM}_{K}(\alpha), \\
q_{n} & \sim \mbox{Categorical}(\boldsymbol{\pi}),  & & n = 1,\dots,N, \label{eq_gen_cat} \\ 
\boldsymbol{\beta}_{k} & \sim \mbox{G}_{0}  & & k = 1,\dots,K, \\
y_{n,t} & \sim \mbox{MNL}( \boldsymbol{\beta}_{k},\boldsymbol{X}_{n,t}), & & n = 1,\dots,N,  \quad t = 1,\dots,T_{n}.
\end{align}

Theoretically, the generative process of the DPM-MNL allows for a large number of populated mixture components. In practice however, the effective number of mixture components, i.e. the number of non-empty components, is considerably smaller than $K$ due to the clustering property of the Dirichlet process. As a reference, Figure \ref{f_lc_mnl} shows the generative process of an LC-MNL model: The generative processes of the DPM-MNL and LC-MNL models resemble each other closely. Yet, in the case of the DPM-MNL model, the truncated stick-breaking process prior on $\boldsymbol{\pi}$ induces shrinkage on the number of effectively populated mixture components, and a prior $\mbox{G}_{0}$ is placed on the component-specific taste parameters $\beta_{1:K}$.

\FloatBarrier
\subsection{Posterior probability}

Next, our goal is to define the posterior probability of the DPM-MNL model parameters $\alpha$ and $\boldsymbol{\beta}_{1:K}$ conditional on the observed choices $\boldsymbol{y} = \left \{ y_{n,1:T_{n}} \right \}_{n=1}^{N}$ and the covariates $\boldsymbol{X} = \{ \boldsymbol{X}_{n,1:T_{n}} \}_{n=1}^{N}$. First, we re-iterate that the component-specific probability distribution functions are MNL models. Hence, the probability of individual $n$ choosing alternative $j$ on choice occasion $t$ conditional on assignment to component $k$ is given by
\begin{equation} \label{eq_mnl}
P(y_{n,t} = j \vert q_{n} = k,\boldsymbol{\beta}_{k}, \boldsymbol{X}_{n,t}) = \frac{\exp ( V(\boldsymbol{X}_{n,t,j}, \boldsymbol{\beta}_{k} )) }{\sum_{j' \in C_{n,t}} \exp ( V (\boldsymbol{X}_{n,t,j'}, \boldsymbol{\beta}_{k} ))},
\end{equation}
where $y_{n,t} \in C_{n,t}$ denotes the observed choice for individual $n$ on occasion $t$. $\boldsymbol{X}_{n,t}$ is a matrix of covariates and $\boldsymbol{X}_{n,t,j}$ is a row of $\boldsymbol{X}_{n,t}$. $\boldsymbol{\beta}_{k}$ is a vector of component-specific taste parameters, and $V(\boldsymbol{X}_{n,t}, \boldsymbol{\beta}_{k})$ is a function giving the deterministic component of utility. $C_{n,t}$ denotes the choice set. (\ref{eq_mnl}) is iterated over alternatives $j \in C_{n,t}$ and choice occasions $t = 1,\ldots,T_{n}$ to obtain the probability of observing choice vector $\boldsymbol{y}_{n} = y_{n,1:T_{n}}$ for individual $n$ conditional on component allocation $q_{n}$:
\begin{equation} \label{eq_pInd_choice}
P(\boldsymbol{y}_{n} \vert q_{n} = k,\boldsymbol{\beta}_{k}, \boldsymbol{X}_{n})
= 
\prod_{t=1}^{T_{n}}
\prod_{j \in C_{n,t}}
P(y_{n,t} = j \vert q_{n} = k, \boldsymbol{\beta}_{k}, \boldsymbol{X}_{n,t})^{[y_{n,t} = j]},
\end{equation} 
where $\boldsymbol{X}_{n} = \boldsymbol{X}_{n,1:T_{n}}$ and $[A]$ is the Kronecker delta, which equals one, if $A$ is true, and zero otherwise. Note that (\ref{eq_pInd_choice}) encapsulates the assumption that an individual's repeated choices are independent from one another conditional on the individual's component assignment. In discrete choice analysis, it is standard practice to adopt this exact conditional independence assumption, when longitudinal choice data are analysed with the help of mixed random utility models \citep[see][]{revelt_mixed_1998}.

Since $q_{n}$ is unobserved, we marginalise over possible values of $q_{n}$ and condition the probability of observing choice vector $\boldsymbol{y}_{n}$ on the prior distribution over components $\boldsymbol{\pi}$ instead of on component assignment $q_{n}$:
\begin{equation} \label{eq_pInd_choice_marg}
P(\boldsymbol{y}_{n} \vert \boldsymbol{\pi}, \boldsymbol{\beta}_{k}, \boldsymbol{X}_{n})
= 
\sum_{k=1}^{K}
\pi_{k}(\boldsymbol{\eta})
\prod_{t=1}^{T_{n}}
\prod_{j \in C_{n,t}}
P(y_{n,t} = j \vert q_{n} = k; \boldsymbol{X}_{n,t},\boldsymbol{\beta}_{k})^{[y_{n,t} = j]}.
\end{equation}
The probability of observing the choice data $\boldsymbol{y} = \boldsymbol{y}_{1:N}$ for the sample is obtained by iterating (\ref{eq_pInd_choice_marg}) over individuals: 
\begin{equation} \label{eq_p_choice}
P(\boldsymbol{y} \vert \boldsymbol{\pi}, \boldsymbol{\beta}_{1:K} ,\boldsymbol{X}) =
\prod_{n=1}^{N}
\sum_{k=1}^{K}
\pi_{k}(\boldsymbol{\eta})
\prod_{t=1}^{T_{n}}
\prod_{j \in C_{n,t}}
P(y_{n,t} = j \vert q_{n} = k; \boldsymbol{X}_{n,t},\boldsymbol{\beta}_{k})^{[y_{n,t} = j]}.
\end{equation}
where $\boldsymbol{X} = \{ \boldsymbol{X}_{n,1:T_{n}} \}_{n=1}^{N}$. Due to the stick-breaking construction of the component weights, $\boldsymbol{\pi}$ is a function of $K-1$ Beta random variables denoted by $\boldsymbol{\eta} = \eta_{1:K-1}$. The joint density of $\boldsymbol{\eta}$ is
\begin{equation} \label{eq_eta_dens}
P(\boldsymbol{\eta} \vert \alpha)  = 
\alpha^{K-1}  \prod_{k=1}^{K-1} (1 - \eta_{k})^{\alpha - 1}.
\end{equation}
Since $\boldsymbol{\eta}$ is also unobserved, we marginalise (\ref{eq_p_choice}) over (\ref{eq_eta_dens}) to obtain the likelihood of the choice data $\boldsymbol{y}$ conditional on the DPM-MNL model parameters $\alpha$ and $\boldsymbol{\beta}$:
\begin{equation} \label{eq_p_choice_marg}
P(\boldsymbol{y} \vert \alpha, \boldsymbol{\beta}_{1:K} ,\boldsymbol{X}) =
\int_{\boldsymbol{\eta}}
\prod_{n=1}^{N}
\sum_{k=1}^{K}
\pi_{k}(\boldsymbol{\eta})
\prod_{t=1}^{T_{n}}
\prod_{j \in C_{n,t}}
P(y_{n,t} = j \vert q_{n} = k; \boldsymbol{X}_{n,t},\boldsymbol{\beta}_{k})^{[y_{n,t} = j]}
P(\boldsymbol{\eta} \vert \alpha) 
d \boldsymbol{\eta},
\end{equation}
where 
\begin{equation}
\pi_{k}(\boldsymbol{\eta}) = 
\begin{cases} 
\eta_{k} \prod_{j=1}^{k-1}(1 - \eta_{j}) & \mbox{ for }  k = 1, \ldots, K - 1 \\
1 - \sum_{k'=1}^{K-1} \pi_{k'} & \mbox{ for }  k = K \\
\end{cases}.
\end{equation}

By Bayes' rule, the posterior probability is equal to the likelihood times the prior divided by the evidence. Hence, 
\begin{equation} \label{eq_bayes}
P(\alpha, \boldsymbol{\beta}_{1:K} \vert \boldsymbol{y}, \boldsymbol{X}) = 
\frac{ P(\boldsymbol{y} \vert \alpha, \boldsymbol{\beta}_{1:K} ,\boldsymbol{X}) f(\alpha) g(\boldsymbol{\beta})}{ \int_{\alpha} \int_{\boldsymbol{\beta}} P(\boldsymbol{y} \vert \alpha, \boldsymbol{\beta}_{1:K} ,\boldsymbol{X}) f(\alpha) g(\boldsymbol{\beta}) d \boldsymbol{\beta} d \alpha} ,
\end{equation}
where $P(\boldsymbol{y} \vert \alpha, \boldsymbol{\beta}_{1:K} ,\boldsymbol{X})$ is defined in (\ref{eq_p_choice_marg}). $f(\alpha)$ denotes the prior density of $\alpha$. $\boldsymbol{\beta}_{k}$, $k = 1, \ldots, K$ are sampled from the base measure $\mbox{G}_{0}$; $g(\boldsymbol{\beta})$ denotes the prior density of $\boldsymbol{\beta}_{1:K}$. The denominator in (\ref{eq_bayes}) represents the model evidence, which does not depend on the DPM-MNL model parameters $\alpha$ and $\boldsymbol{\beta}$.

\section{Model inference} \label{S_modelInf}

\subsection{Overview}

Having defined the posterior probability of the DPM-MNL model parameters, we wish to devise a method for posterior inference. Two factors complicate inference in the DPM-MNL model: First, exact inference is not possible, as both the numerator and the denominator of the posterior probability (\ref{eq_bayes}) involve integrations that are not analytically tractable. Second, the likelihood function (\ref{eq_p_choice_marg}) involves a summation of the $K$ component-specific MNL kernels. Hence, the likelihood function is likely to exhibit multiple modes and a closed-form expression for the gradient of the log-likelihood function does not exist.

In general, inference in Dirichlet process mixture models has been carried out using Markov Chain Monte Carlo (MCMC) methods \citep{neal_markov_2000,ishwaran_gibbs_2001} and variational inference methods \citep{blei_variational_2006}. MCMC methods treat the latent component assignments as random model parameters and approximate a posterior distribution through samples from a Markov Chain, whose stationary distribution is the posterior distribution of interest \citep[e.g.][]{blei_variational_2006}. Variational inference methods hinge on finding a variational distribution over the latent variables to approximate the difficult-to-compute posterior distribution \citep[e.g.][]{wainwright_graphical_2008}. The variational distribution is characterised by its own variational parameters, which are chosen such that the variational distribution and the posterior of interest are close to one another. Subsequently, inference on the model parameters proceeds with the variational distribution as a surrogate posterior distribution. 

Both inference approaches have been successfully employed in numerous empirical applications \citep[see][for surveys]{carvalho_particle_2010,wang_fast_2011} but are subject to limitations: MCMC methods are computationally intensive and do not scale well to larger datasets (for an argumentation in the context of Dirichlet process mixture models, see \citeauthor{wang_fast_2011}, \citeyear{wang_fast_2011}; for an argumentation in the context of discrete choice models, see \citeauthor{braun_variational_2010}, \citeyear{braun_variational_2010}). In particular, when models depend on discrete latent quantities---such as labels in a mixture model---, MCMC samplers may exhibit poor mixing, and label-switching issues need to be addressed \citep[e.g.][]{gelman_bayesian_2013}. Variational inference methods allow for fast inference but give estimates that may not be asymptotically efficient. Thus, variational inference methods are most suitable for repeated inference on very large datasets and for applications where precise parameter estimates are not a primary concern \citep{blei_variational_2017}.

In this paper, we conceive inference in a Dirichlet process mixture model as a missing data problem and leverage the expectation maximisation (EM) algorithm \citep{dempster_maximum_1977,mclachlan_em_2008} for maximum a posteriori (MAP) estimation of the DPM-MNL model parameters. MAP estimation is a computationally efficient Bayesian inference approach, whose objective is to identify the mode of the log-posterior density. From a practical point-of-view, MAP estimation extends maximum-likelihood estimation by accounting for prior information about the distribution of the model parameters. The implementation of MAP estimation approaches is generally possible via gradient-based optimisation routines. However, in the case of the proposed DPM-MNL model, gradient-based optimisation approaches are not feasible due to the complications outlined in the first paragraph of this subsection. 

In principle, an EM algorithm for MAP estimation problems alternates between an expectation step (E-step) and a maximisation step (M-step) until a convergence criterion is satisfied. In the E-step, the expectation of the posterior density is computed. The objective of the subsequent M-step is to identify the mode of the expected posterior density by maximising over the set of unknown model parameters. Our main rationale for leveraging the EM algorithm for posterior inference in the DPM-MNL model is that the M-step results in a set of standard optimisation problems that are much simpler than the direct optimisation of the posterior density. In addition, label-switching is not a concern due to the deterministic nature of the EM algorithm. For general discussions of the properties of the EM algorithm, we refer to the literature \citep{dempster_maximum_1977,mclachlan_em_2008,train_discrete_2009}.

The EM algorithm is a fast, scalable and precise inference approach, which is suitable for the estimation of a variety of latent variable models \citep[see e.g.][]{mclachlan_em_2008}. In the domain of discrete choice analysis, \citet{bhat_endogenous_1997} presents an EM algorithm for the estimation of an LC-MNL model. \citet{train_em_2008,train_discrete_2009} devises EM algorithms for the estimation of M-MNL models with a variety of nonparametric and semi-nonparametric mixing distributions. Moreover, \citet{vij_random_2017} introduce an EM algorithm for inference in a M-MNL model with a gridded mixing distribution. \citet{sohn_expectation-maximization_2016} derives an EM algorithm for the integrated choice and latent variable model \citep{walker_generalized_2002}. 

\subsection{Derivation of an expectation maximisation algorithm}

We begin the derivation of the EM algorithm for the DPM-MNL model by writing out the joint distribution of the DPM-MNL model parameters $\alpha$, $\boldsymbol{\beta}_{1:K}$ and the latent variables $q_{1:N}$: 
\begin{equation}  \label{eq_joint}
\begin{split}  
P (\alpha, \boldsymbol{\beta}_{1:K}, q_{1:N}, \boldsymbol{y},  \boldsymbol{X})  = &
\left (
\frac{\Gamma(N + 1)}{\Gamma(1 + \sum_{n=1}^{N} [q_{n} = K])}
\alpha^{K-1} 
\prod_{k=1}^{K-1} 
\frac{ \Gamma(\alpha + v_{k+1})}{\Gamma(1 + \alpha + v_{k+1})}
\right ) \\
& \left (
\prod_{n=1}^{N}
\prod_{t=1}^{T_{n}} 
\prod_{k=1}^{K}
\prod_{j \in C_{n,t}} 
P(y_{n,t} = j \vert q_{n} = k; \boldsymbol{X}_{n,t},\boldsymbol{\beta}_{k})^{[y_{n,t} = j] \cdot [q_{n} = k] }
\right ) \\
& f(\alpha)\\
& \prod_{k=1}^{K}
g(\boldsymbol{\beta}_{k}), \\
\end{split}
\end{equation}
where $v_{k} = \sum_{k'=k}^{K} \sum_{n=1}^{N} [q_{n} = k']$. The joint probability (\ref{eq_joint}) consists of four factors (one in each line): The first factor follows from the fact that the distribution of the latent variables $q_{1:N}$ is the generalised-Dirichlet-multinomial distribution. This is because $N$ samples from a categorical distribution define a multinomial distribution with $N$ trials, and the multinomial distribution is conjugate to the GEM distribution, a special case of the generalised Dirichlet distribution. The second factor factorises the component-specific MNL kernels for each component, each choice occasion and each decision-maker. The third factor represents the prior density of $\alpha$, and the fourth factor factorises the prior densities of the taste vectors $\boldsymbol{\beta}_{1:K}$. 

By Bayes' rule the joint probability (\ref{eq_joint}) is proportional to the posterior density of interest, i.e. $P(\alpha, \boldsymbol{\beta}_{1:K} \vert \boldsymbol{y}, \boldsymbol{X}) \propto P(\alpha, \boldsymbol{\beta}_{1:K}, q_{1:N}, \boldsymbol{y},  \boldsymbol{X})$, where the normalising constant, i.e. the marginal likelihood in the denominator, can be disregarded as it does not depend on the DPM-MNL model parameters. Finding the mode of the posterior density is therefore equivalent to finding the mode of the joint probability. The EM algorithm facilitates the optimisation of the joint probability by imputing the missing data in the E-step. In the sequel, we present the E- and M-steps of the algorithm and assume that the algorithm is initialised with a set of starting values $\{ \alpha^{(\tau)}, \boldsymbol{\beta}^{(\tau)} \}$.

\paragraph{E-step}
In the E-step, we compute the expectation $Q (\alpha, \boldsymbol{\beta} \vert \alpha^{(\tau)}, \boldsymbol{\beta}^{(\tau)})$ of the complete-data posterior density with respect to the latent variables conditional on the current parameter estimates. In the case of (\ref{eq_joint}), the probability mass functions $[q_{n} = k]$ for $n = 1,\ldots,N$, $k = 1,\ldots,K$ are the sufficient statistics required for the estimation of the unknown model parameters. Hence, we can determine the expectation of the complete-data posterior density by computing the expectation of each sufficient statistic $[q_{n} = k]$ conditional on the current parameter estimates $\alpha^{(\tau)}$, $\boldsymbol{\beta}^{(\tau)}$. From Bayes' rule, we have
\begin{equation} \label{eq_eStep}
\begin{split}
\omega_{n,k} 
& \equiv \mathbf{E} \left ([q_{n} = k] \vert \boldsymbol{y}_{n}; \alpha^{(\tau)}, \boldsymbol{\beta}^{(\tau)}\right ) \\
& = \frac{P (q_{n} = k \vert \alpha^{(\tau)}) P(\boldsymbol{y}_{n} \vert q_{n} = k; \boldsymbol{X}_{n},\boldsymbol{\beta}_{k}^{(\tau)})}{\sum_{k'=1}^{K} P(q_{n} = k' \vert \alpha^{(\tau)} ) P(\boldsymbol{y}_{n} \vert q_{n} = k'; \boldsymbol{X}_{n},\boldsymbol{\beta}_{k'}^{(\tau)})}, \\
& n = 1,\ldots,N, \quad k = 1,\ldots,K,
\end{split} 
\end{equation}
where 
\begin{equation}
P(q_{n} = k \vert \alpha^{(\tau)}) = 
\begin{cases} 
\frac{\left ( \alpha^{(\tau)} \right )^{k-1}}{\left ( 1 + \alpha^{(\tau)} \right )^{k}} & \mbox{ for }  k = 1, \ldots, K - 1 \\
1 - \sum_{k'=1}^{K-1} P(q_{n} = k' \vert \alpha^{(\tau)}) & \mbox{ for }  k = K \\
\end{cases}.
\end{equation}
Consequently, the expectation of the complete-data posterior density is
\begin{equation} \label{eq_complete_post}
\begin{split}  
Q (\alpha, \boldsymbol{\beta} \vert \alpha^{(\tau)}, \boldsymbol{\beta}^{(\tau)})  = &
\left (
\frac{\Gamma(N + 1)}{\Gamma(1 + \sum_{n=1}^{N} [q_{n} = K])}
\alpha^{K-1} 
\prod_{k=1}^{K-1} 
\frac{ \Gamma(\alpha + v_{k+1})}{\Gamma(1 + \alpha + v_{k+1})}
\right ) \\
& \left (
\prod_{n=1}^{N}
\prod_{t=1}^{T_{n}} 
\prod_{k=1}^{K}
\prod_{j \in C_{n,t}} 
P(y_{n,t} = j \vert q_{n} = k; \boldsymbol{X}_{n,t},\boldsymbol{\beta}_{k})^{[y_{n,t} = j] \cdot \omega_{n,k} }
\right )\\
& f(\alpha)\\
& \prod_{k=1}^{K}
g(\boldsymbol{\beta}_{k}), \\
\end{split}
\end{equation}
where $v_{k} = \sum_{k'=k}^{K} \sum_{n=1}^{N} \omega_{n,k'}$.

\paragraph{M-step}
In the M-step, we identify the mode of the surrogate function (\ref{eq_complete_post}) by maximising over the set of unknown model parameters. We update the model parameters by solving
\begin{equation} \label{eq_mStep}
\{ \alpha^{(\tau+1)}, \boldsymbol{\beta}^{(\tau+1)} \}  =  
\operatorname*{arg\,max}_{\{ \alpha, \boldsymbol{ \beta} \}}
\ln Q (\alpha, \boldsymbol{\beta} \vert \alpha^{(\tau)}, \boldsymbol{\beta}^{(\tau)}),
\end{equation}
which can be separated into into two comparatively easy optimisation problems. First, the concentration parameter is updated by numerically solving:
\begin{equation} \label{eq_mStep1}
\begin{split}
\alpha^{(\tau+1)}  = & \operatorname*{arg\,max}_{\alpha}
\Bigg \{ 
(K - 1) \ln \alpha + 
\sum_{k=1}^{K-1} \ln \Gamma(\alpha + w_{k+1}) - \\
& \sum_{k=1}^{K-1} \ln \Gamma(1 + \alpha + w_{k}) + \ln f (\alpha)
\Bigg \},
\end{split}
\end{equation}
where $w_{k} = \sum_{k'=k}^{K} \sum_{n=1}^{N} \omega_{n,k'}$. Note that the last summand of the objective function represents the log prior density of $\alpha$. Second, the component-specific parameters are updated:
\begin{equation} \label{eq_mStep2}
\begin{split}
\boldsymbol{\beta}_{k}^{(\tau+1)} = & \operatorname*{arg\,max}_{\boldsymbol{\beta}_{k}}  
\Bigg \{
\sum_{n=1}^{N} 
\sum_{t=1}^{T_{n}} 
\sum_{j \in C_{n,t}} 
\omega_{n,k} \cdot [y_{n,t} = j] \cdot  \ln P \left (y_{n,t} = 1 \vert q_{n} = k; \boldsymbol{X}_{n,t},\boldsymbol{\beta}_{k} \right ) + \\
& \ln g(\boldsymbol{\beta}_{k})
\Bigg \},
\quad k = 1, \ldots, K,
\end{split} 
\end{equation}
where the first summand of the objective function is essentially the log-likelihood of a weighted multinomial logit model. The second summand of the objective function represents the prior density over the taste vector $\boldsymbol{\beta}_{k}$. If the logarithm of the prior density and the gradient of the logarithm of the prior density exist in closed form, (\ref{eq_mStep2}) can be solved with the help of standard gradient-based optimisation routines.

\subsection{Practicalities} \label{S_practicalities}

Parameter estimates for the DPM-MNL model are obtained by cycling through the E- and M-steps presented above, until a convergence criterion is satisfied. As each iteration of the EM algorithm results in an improvement of the expected value of the log-posterior density, convergence can be assessed by considering the improvement of the expected log-posterior density relative to the previous iteration. In the subsequent applications of the inference approach, the execution of the EM algorithm is terminated, if the improvement of the expected log-posterior density between successive iterations is less than 0.01\% of the current expected log-posterior density. 

The EM algorithm is a deterministic algorithm, which, given a certain set of starting values will always converge to the same local optimum. However, the algorithm does not guarantee convergence to a global optimum. In practice, it is therefore critical to choose good starting values to avoid that the algorithm terminates in a local optimum. The analyst may use her intuition or may employ a systematic approach to choose starting values. In the subsequent applications of the EM algorithm, we adopt \citeauthor{train_em_2008}'s (\citeyear{train_em_2008}) procedure, i.e. we randomly partition the sample into $K$ groups and estimate separate MNL models for each of the groups. The MNL estimates are then used as starting values for the component-specific taste parameters and mixture components are assigned equal weights. 

Moreover, the truncation level $K$ of the truncated stick-breaking process prior must be set by the analyst. Generally, the choice of $K$ affects the quality of the approximation of the Dirichlet process via the truncated stick-breaking process, but also impacts the computational tractability of the inference approach \citep{ohlssen_flexible_2007}. In that vein, the literature has considered different truncation levels: For example, \citet{ishwaran_gibbs_2001} as well as \citet{li_bayesian_2013} use $K = 150$, while \citet{blei_variational_2006} employ a truncation level of  $K = 20$; \citet{gelman_bayesian_2013} suggest that $K \in [20;50]$ suffices for most practical applications. In general, $K$ is naturally truncated by the sample size. In addition, the analyst may have a strong prior belief about the number of heterogeneity components that are required to explain the observed data and may set $K$ based on her prior belief. Either way, $K$ should be set such that components with a large indices have low probabilities of being occupied \citep{gelman_bayesian_2013}. To this end, it is advisable to place an informative prior on the concentration parameter $\alpha$, since $\alpha$ has an immediate effect on the number of occupied mixture components \citep{ohlssen_flexible_2007}. In the subsequent applications of the inference approach, we follow \citet{li_bayesian_2013} and set $K = 150$ to assure a close approximation of the Dirichlet process. In addition, we follow \citet{ishwaran_approximate_2002} and let $\alpha \sim \mbox{Gamma}(2,2)$, i.e. the prior density of $\alpha$ is Gamma with shape 2 and scale 2. 

Thus far, we have not explicitly specified the base measure $\mbox{G}_{0}$. To facilitate model inference, it is generally useful to let $\mbox{G}_{0}$ be conjugate to the kernel \citep{gelman_bayesian_2013}. However, in the case of the proposed DPM-MNL model, conjugacy is not a desideratum, as the MNL kernel does not have a conjugate prior. Instead, we would like to employ a base measure where the log-density and the gradient of the log-density can be expressed in closed form to allow for the application of standard gradient-based optimisation routines. \citet{gelman_bayesian_2013} advise against the use of diffuse base measures, as a high variance of the base measure may penalise the addition of new mixture components and may thus limit the flexibility of the mixing distribution. An obvious choice for $\mbox{G}_{0}$ is the normal distribution. In the subsequent applications of the DPM-MNL model, we let $\mbox{G}_{0} = \mbox{N}(0,5^{2})$. For taste parameters that are constrained to be strictly positive or negative---such as the taste parameter capturing sensitivity to cost---, we employ half-normal priors with scale 5. For priors to be meaningful, the scale of the corresponding parameters may have to be adjusted. In the subsequent applications of the DPM-MNL model, we scale the covariates such that the absolute values of the coefficient estimates of a standard MNL model applied to the same dataset as the DPM-MNL model are between 0.1 and 1. 

\section{Simulation study} \label{S_simStudy}

In this section, we present a simulation study consisting of four Monte Carlo experiments to demonstrate the behaviour of the proposed DPM-MNL model for different taste parameter distributions. 

\subsection{Synthetic data generating process}

For the simulation study, we generate multiple synthetic samples, each comprising 2,000 individuals, who are pseudo-observed to each complete eight choice tasks. The choice scenarios include three unlabelled alternatives, which are characterised by three attributes, namely in-vehicle travel time, out-of-vehicle travel time and travel cost. Table \ref{table_simStudy02_covariates} in the appendix details the data generating process of the simulated attribute levels.  

The individuals are assumed to be utility maximisers and to evaluate alternatives based on the following utility specification:
\begin{equation}
U_{n,t,j} =  \left ( \mbox{ivtt}_{n,t,j} \beta_{n,\mbox{ivtt}} + \mbox{ovtt}_{n,t,j} \beta_{n,\mbox{ovtt}}+ \mbox{cost}_{n,t,j} \right ) \beta_{\mbox{n,cost}} + \epsilon_{n,t,j},
\end{equation}
where $n$ indexes individuals, $t$ indexes choice occasions and $j$ indexes alternatives. $\beta_{n,\mbox{ivtt}}$, $\beta_{n,\mbox{ovtt}}$ and $\beta_{n,\mbox{cost}}$ represent taste parameters, which respectively pertain to in-vehicle travel time (ivtt), out-of-vehicle travel time (ovtt) and travel cost. $\epsilon_{n,t,j}$ is a disturbance assumed to be i.i.d. $\mbox{Gumbel} \left( 0,\frac{\pi^2}{6} \right)$. $\beta_{n,\mbox{ivtt}}$, $\beta_{n,\mbox{ovtt}}$ and $\beta_{n,\mbox{cost}}$ vary randomly across individuals. The utility function is specified in willingness-to-pay space so that $\beta_{n,\mbox{ivtt}}$ and $\beta_{n,\mbox{ovtt}}$ are identical to the implicit values of the corresponding attributes. 

In each of the four Monte Carlo experiments, the distributions of $\beta_{n,\mbox{ivtt}}$, $\beta_{n,\mbox{ovtt}}$ and $\beta_{n,\mbox{cost}}$ are manipulated: In the first experiment, $\beta_{n,\mbox{ivtt}}$ and $\beta_{n,\mbox{ovtt}}$ are sampled from a bivariate normal distribution so that the joint distribution of the two parameters is uni-modal with light tails. In the second experiment, $\beta_{n,\mbox{ovtt}}$ is log-normally distributed, while $\beta_{n,\mbox{ivtt}}$ is normally distributed; hence, the joint distribution of the two parameters is uni-modal and exhibits a heavy tail for one of the marginals. In the third experiment, $\beta_{n,\mbox{ivtt}}$ and $\beta_{n,\mbox{ovtt}}$ are sampled from a two-component-mixture of normals so that the joint distribution of the two parameters is bi-modal. In the fourth experiment, the joint distribution of $\beta_{n,\mbox{ivtt}}$ and $\beta_{n,\mbox{ovtt}}$ is tri-modal, as the two parameters are drawn from a three-component mixture of normals. In each of the four experiments, the negative of $\beta_{n,\mbox{cost}}$ is log-normally distributed to assure strict negativity of $\beta_{n,\mbox{cost}}$. In each of the four experiments, the location parameter of the distribution of $\beta_{n,\mbox{cost}}$ is set such that the error rate is roughly 7\%, i.e. in 7\% of the cases, decision-makers deviate from the deterministically best alternative due to the stochastic component. Table \ref{table_simStudy02_tastes} in the appendix gives the data generating processes of the taste parameters for the four Monte Carlo experiments.

\subsection{Method}

For each of the four Monte Carlo experiments, we estimate the proposed DPM-MNL model, using the inference approach presented in Section \ref{S_modelInf}. We implement the inference method by writing our own MATLAB code and use Train's (\citeyear{train_em_2008}) procedure to obtain starting values. 

\subsection{Results}

Figure \ref{figure_sim1} visualises the true and the estimated taste parameter distributions for the simulation study. Each subfigure (see Figures \ref{figure_sim1a}--\ref{figure_sim1d}) corresponds to one of the four Monte Carlo experiments. In each subfigure, the first row shows histograms of the true joint density of $\beta_{n,\mbox{ivtt}}$ and $\beta_{n,\mbox{ovtt}}$; the second figure in the first row of each subfigure additionally displays the estimated point masses of the DPM-MNL model. By design, the DPM-MNL model yields discrete, non-smooth heterogeneity representations. To allow for a comparison of the true and the estimated distribution function, we estimate kernel density functions with normal kernel and bandwidth 2.5 for both the true and the estimated taste parameter distributions.\footnote{In the case of the DPM-MNL model, the kernel density functions are estimated based on 2,000 random draws from the estimated taste parameter distributions.} Plots of these kernel density functions are given in the second row of each subfigure. 

Overall, we observe that the DPM-MNL performs well at recovering differently-shaped taste parameter distributions. The DPM-MNL model captures unobserved taste heterogeneity by parsimoniously placing few mass points in the hypothesis space. The estimated locations of these mass points may not necessarily coincide with the modes of the true heterogeneity distributions. However, the estimated kernel density functions show that the DPM-MNL model is able to correctly identify the effective support and the modes of the true taste parameters distributions in all four Monte Carlo experiments. 

The estimated concentration parameters of the $\mbox{GEM}_{K}$ distribution are 6.9, 7.0, 7.5 and, respectively, 7.5 for each of the four Monte Carlo experiments. Given these estimates of $\alpha$, the expected numbers of mixture components with an occupancy of at least one decision-maker are 41, 42, 46 and 44, respectively, in each of the four Monte Carlo experiments. 

\begin{figure}[p]
\centering

	\begin{subfigure}[a]{\textwidth}
	\includegraphics[width = \textwidth]{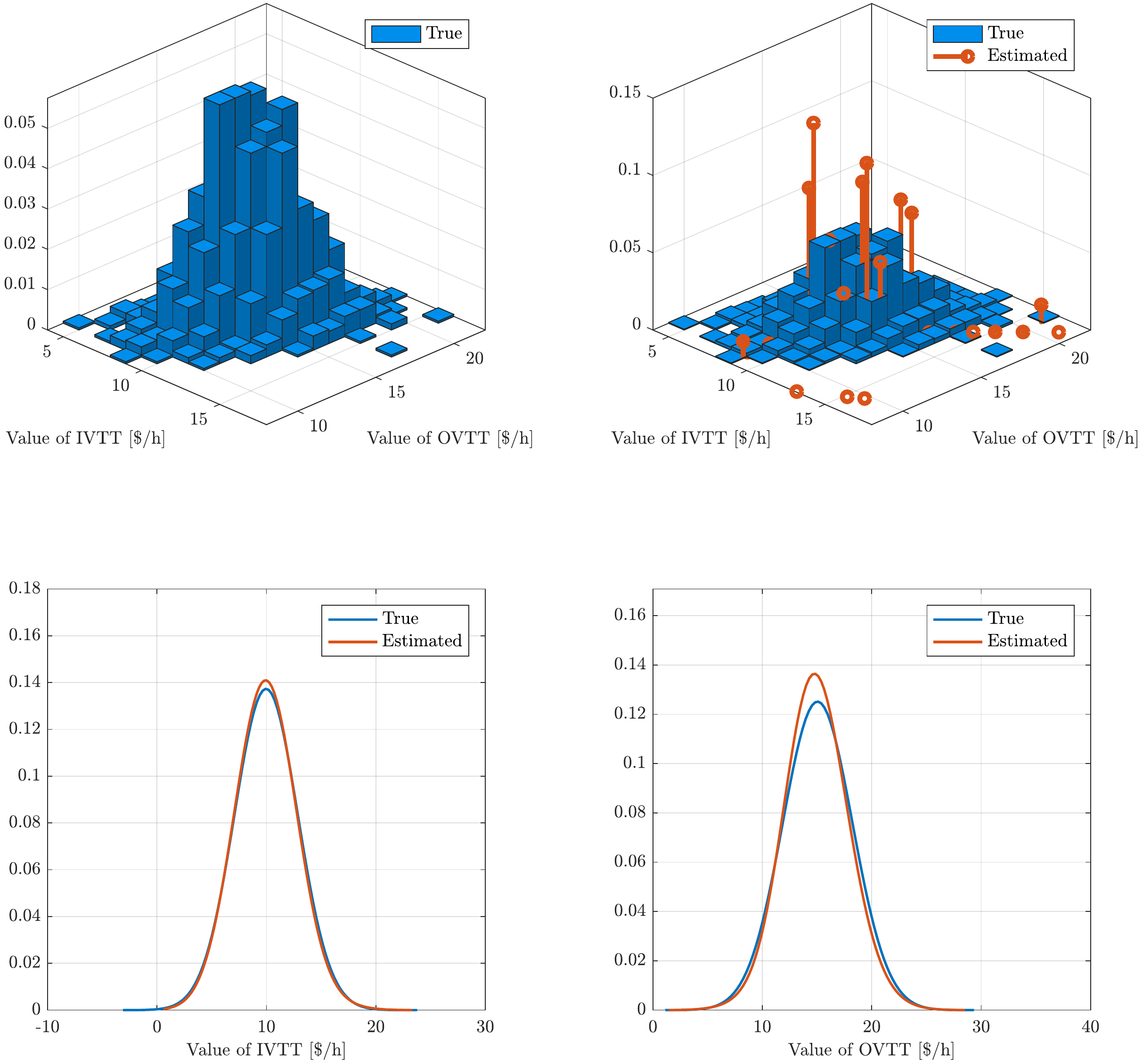}
	\caption{Experiment I: Uni-modal distribution} \label{figure_sim1a}
	\end{subfigure}
	
\end{figure}
\begin{figure}[h]
\ContinuedFloat
	
	\begin{subfigure}[b]{\textwidth}
	\includegraphics[width = \textwidth]{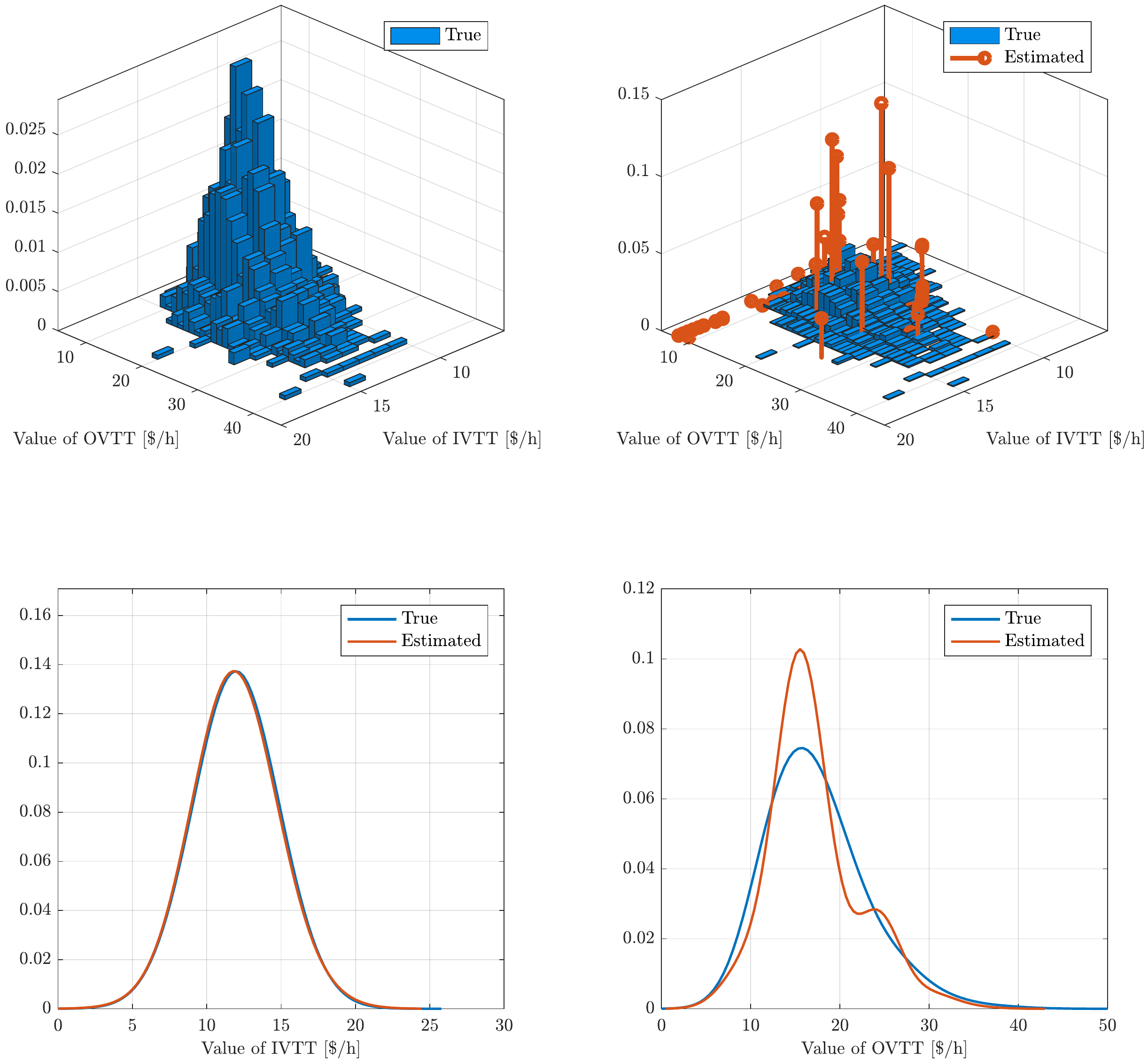}
	\caption{Experiment II: Uni-modal distribution with one heavy-tailed marginal} \label{figure_sim1b}
	\end{subfigure}
	
\end{figure}
\begin{figure}[h]
\ContinuedFloat
	
	\begin{subfigure}[c]{\textwidth}
	\includegraphics[width = \textwidth]{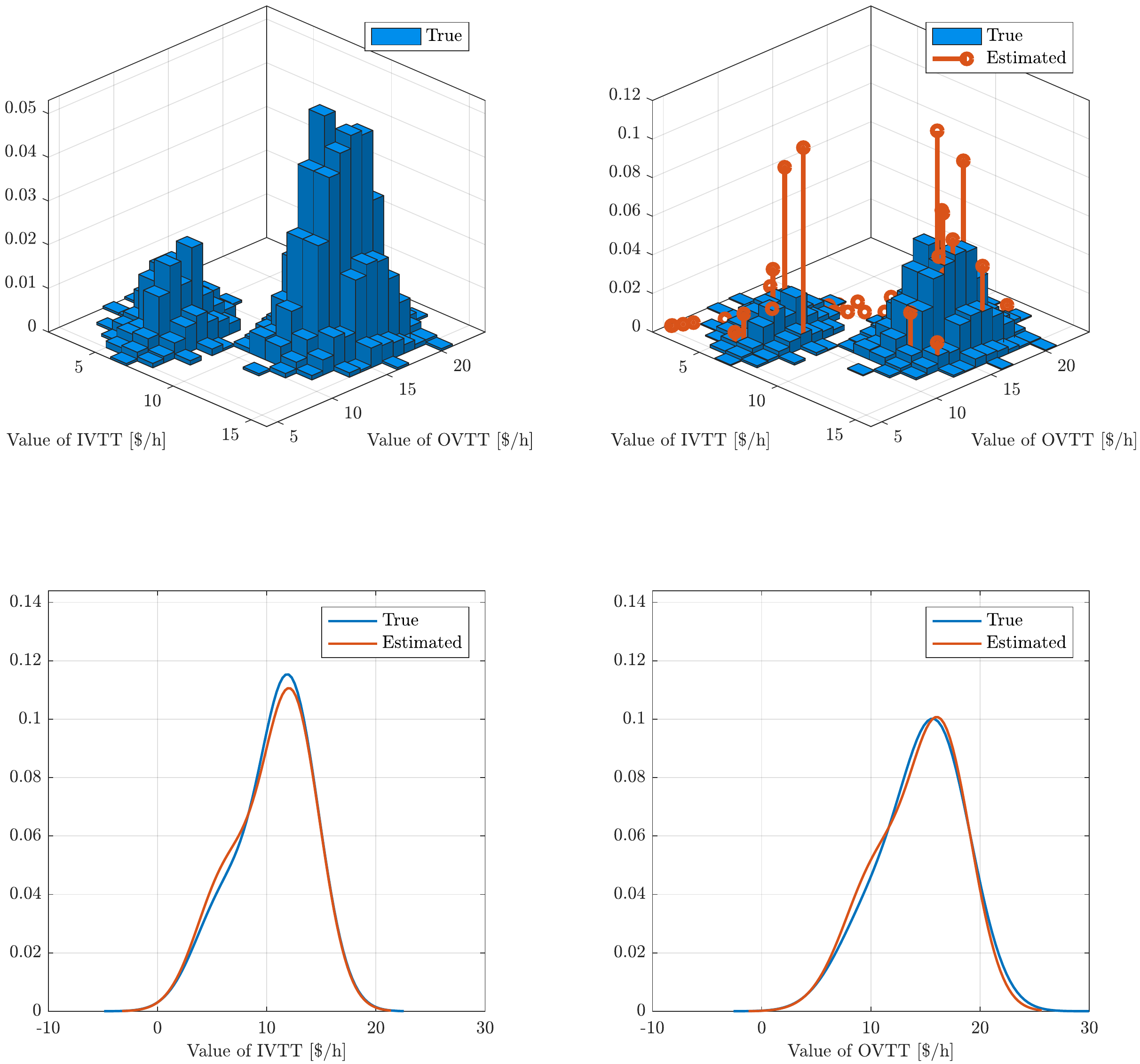}
	\caption{Experiment III: Bi-modal distribution} \label{figure_sim1c}
	\end{subfigure}
	
\end{figure}
\begin{figure}[h]
\ContinuedFloat
	
	\begin{subfigure}[d]{\textwidth}
	\includegraphics[width = \textwidth]{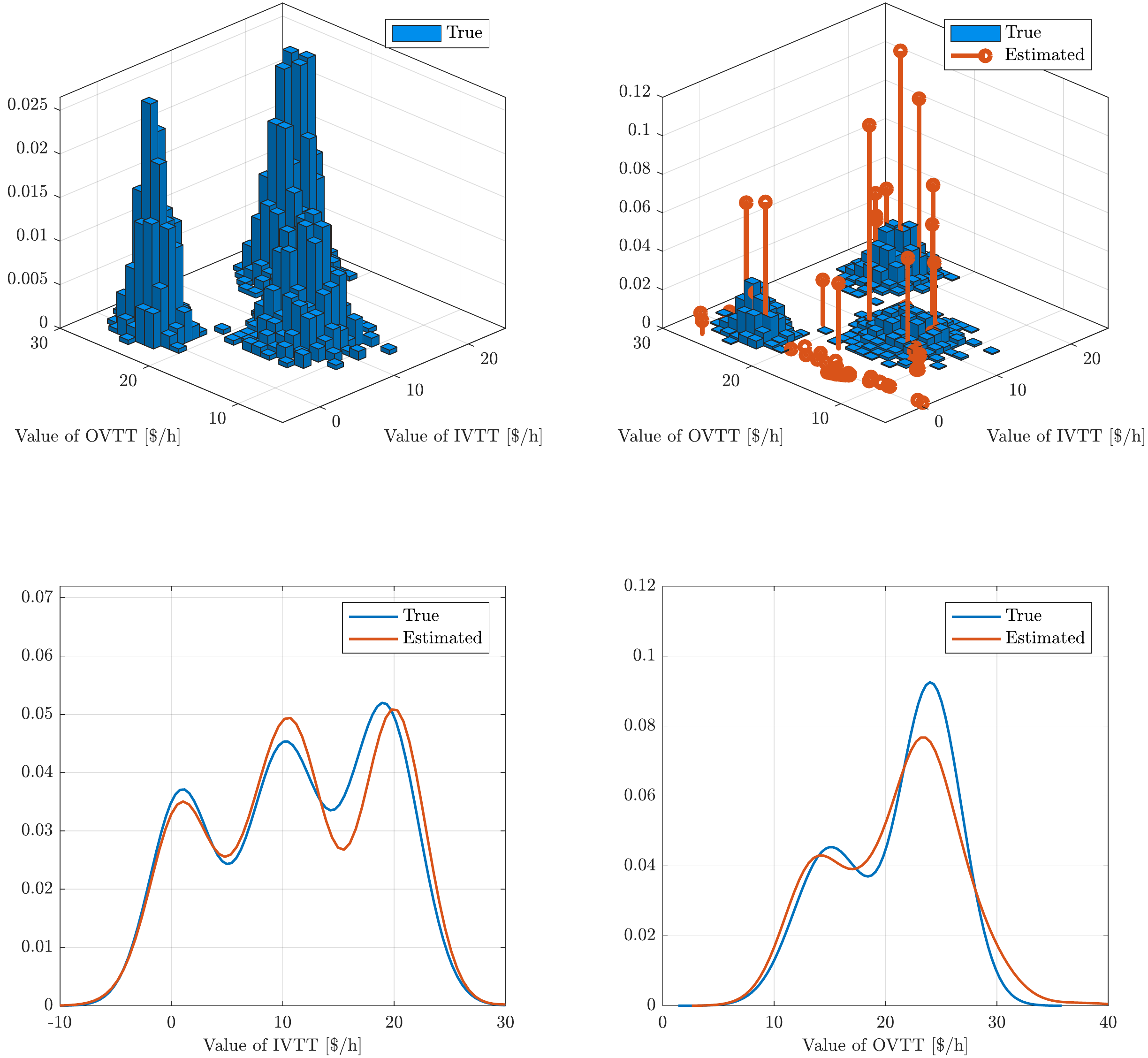}
	\caption{Experiment IV: Tri-modal distribution} \label{figure_sim1d}
	\end{subfigure}
	
\caption{True and estimated taste parameter distributions for the simulation study} \label{figure_sim1}
\end{figure}

\section{Case study} \label{S_caseStudy}

In this section, we empirically validate the proposed model framework in a case study on motorists' route choice preferences.  

\subsection{Data}

Data for our analysis are sourced from the German Value of Time and Reliability Study, which was commissioned by the German Federal Ministry of Transport and Digital Infrastructure to obtain estimates of travellers' valuation of travel time and reliability for the Federal Transport Investment Plan 2030, a strategic programme for the appraisal of federal transport infrastructure projects in Germany. The German Value of Time and Reliability Study involved a series of stated choice experiments to elicit tastes with respect to strategic level of service attributes in the context of mode choice, route choice, travel itinerary choice, workplace choice and residential location choice. More information about the scope of the study and the data collection is provided by \citet{axhausen_ermittlung_2015} and \citet{ehreke_experiences_2014}. For the case study, we consider a stated choice experiment on motorists' route choice preferences. The choice tasks required respondents to choose the best of two route alternatives, each of which was characterised by five attributes, namely free-flow travel time, access time, time spent in congested traffic conditions, probability of a significant delay and travel cost. The sample considered for the case study comprises 3,579 cases and 455 individuals. 

\subsection{Method}

The stated choice data are used to estimate the proposed DPM-MNL model. In addition, we estimate MNL, PM-MNL and LC-MNL models to benchmark the performance of the proposed DPM-MNL model against established modelling approaches. The DPM-MNL model is estimated, using the inference approach presented in Section \ref{S_modelInf}; we write our own MATLAB code to implement the inference method and employ Train's (\citeyear{train_em_2008}) method to obtain starting values. The PM-MNL models are estimated, using maximum-simulated likelihood methods \citep{train_discrete_2009} in conjunction with PythonBiogeme \citep{bierlaire_pythonbiogeme:_2016}. For each individual, 2,000 simulation draws generated via the Modified Latin Hypercube Sampling method \citep{hess_use_2006} are used. For the estimation of the LC-MNL models, we employ the EM algorithm as outlined in \citet{train_em_2008} and write our own MATLAB code to implement the inference approach. Again, Train's (\citeyear{train_em_2008}) procedure is employed to obtain starting values.

Two different PM-MNL models are considered: The first model assumes that the implicit attribute values are normally distributed, while the second model assumes log-normally distributed implicit attributes values. In both models, the taste parameter capturing sensitivity to cost is assumed to be log-normally distributed. In the first model, utility is specified in willingness-to-pay space to assure that the implicit attribute values follow the desired random distribution. In the second model, it is not imperative to define the utility function in willingness-to-pay space, as the ratio of two log-normally distributed random variables is again log-normally distributed. Therefore as well as for numerical reasons, utility in the second model is specified in preference space. In both model specifications, the random taste parameters are assumed to be independent from one another. 

In the case of the LC-MNL model, we are required to estimate multiple model specifications with varying numbers of mixture components, since model complexity represents an exogenous model parameter in a LC-MNL model. A final model specification can be selected based on a consideration of statistical information criteria such as the Akaike Information Criterion \citep[AIC; ][]{akaike_new_1974} and the Bayesian Information Criterion \citep[BIC; ][]{schwarz_estimating_1978}. In that vein, we initiate the specification search by first estimating a two-component LC-MNL model and increment the number of mixture components in subsequent estimation runs. A comparison of the estimated LC-MNL model specifications is given in Table \ref{table_caseStudy_lc_ic}. In general, AIC and BIC can attain their minima at different model specifications, because the BIC penalises model complexity more strictly than the AIC. In the present application, the BIC suggests that a model specification with six latent classes is optimal, while the AIC suggests that a model specification with 14 latent classes should be preferred. 

\begin{table}[h]
\small
\input{table_caseStudy_lc_ic.tex}
\caption{Comparison of LC-MNL model specifications (N = 455)} \label{table_caseStudy_lc_ic}
\end{table}
\FloatBarrier

The performance of the DPM-MNL, MNL, PM-MNL and LC-MNL models is assessed in terms of both in-sample fit and out-of-sample predictive ability. To evaluate the out-of-sample predictive ability of each of the considered models, we employ a ten-fold cross-validation approach. To this end, the sample is randomly divided into ten subsamples of equal size. In each rotation of the validation procedure, each of the considered models is trained on nine of the subsets, while the held-out subsample is used to compute the predictive log-likelihood for the fold. In each of the ten rotations of the cross-validation procedure, a different subsample is held-out for validation so that in the end, each of the ten subsamples is used once for validation. The average of the predictive log-likelihood values for each of the folds gives the ten-fold cross-validated log-likelihood.

\subsection{Results}

Table \ref{table_caseStudy_comparison} gives log-likelihood values indicating the in-sample fit and the out-of-sample predictive ability for each of the estimated models. The hypothesis of taste homogeneity can be soundly rejected, as the MNL model is substantially outperformed by the competing models. Moreover, the PM-MNL models with discrete heterogeneity representations outperform the two M-MNL with continuous parametric mixing distribution in terms of both in-sample fit and out-of-sample predictive ability. The DPM-MNL provides the best in-sample fit and out-of-sample predictive ability of all considered models. We re-iterate that inference in the LC-MNL model required an extensive specification search (see Table \ref{table_caseStudy_lc_ic}), whereas inference in the DPM-MNL model was instantaneous. 

\begin{table}[h]
\small
\input{table_caseStudy_comparison.tex}
\caption{Model comparison} \label{table_caseStudy_comparison}
\end{table}

\begin{figure}[h]
\centering
\includegraphics[width = \textwidth]{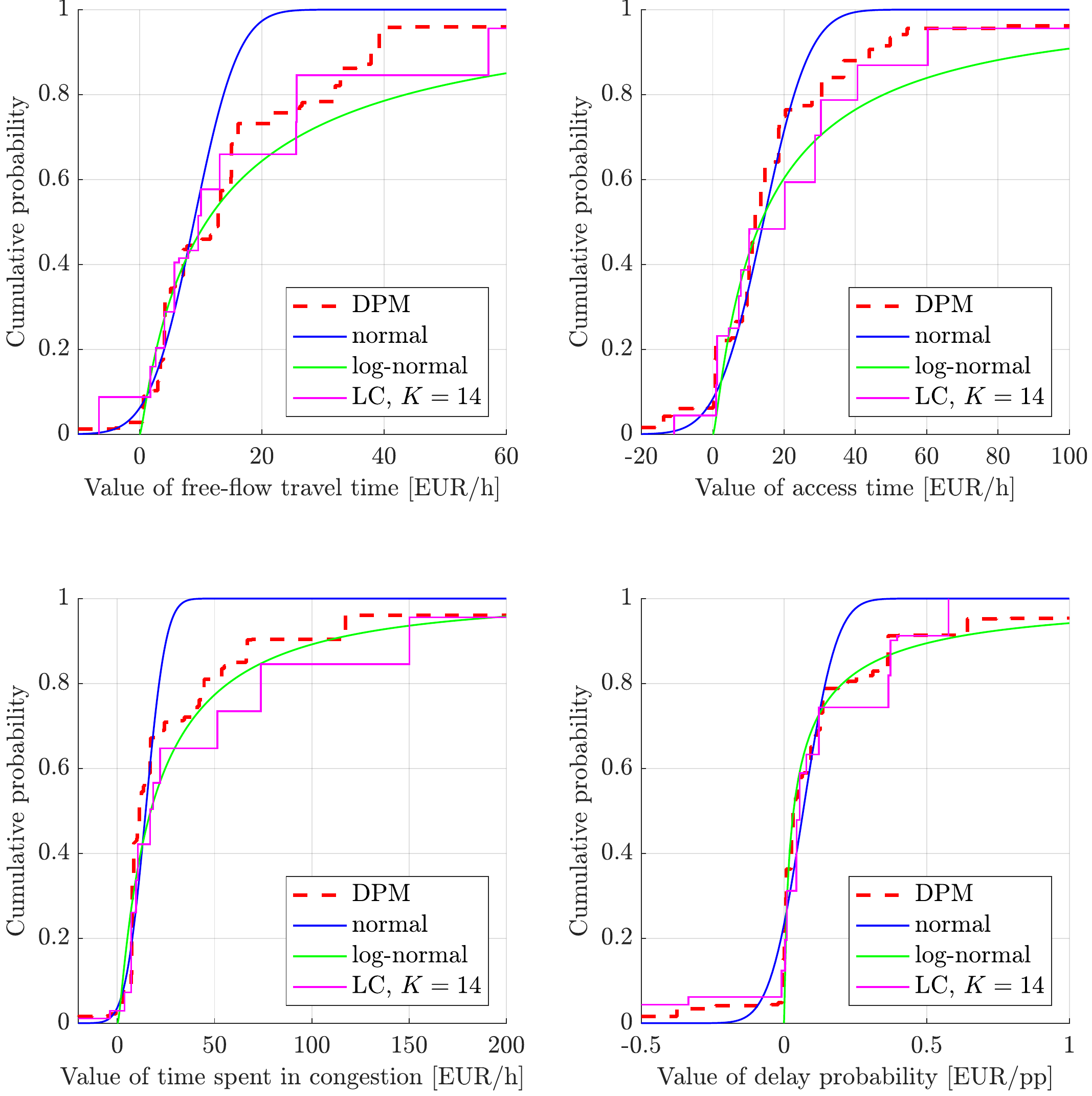}
\caption{Estimated cumulative distribution functions of implicit attribute values} \label{figure_caseStudy1}
\end{figure}

\begin{figure}[h]
\centering
\includegraphics[width = 0.5 \textwidth]{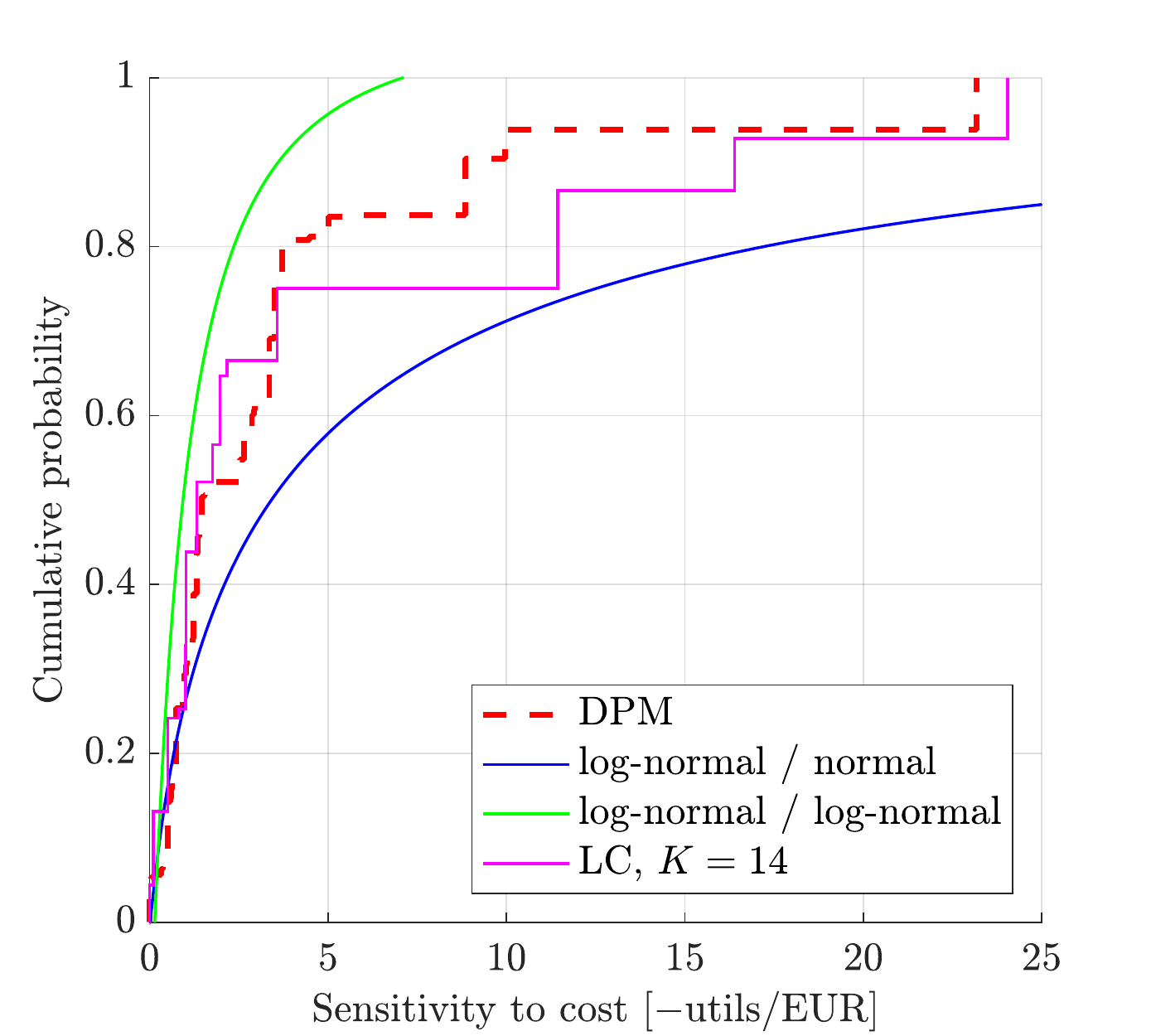}
\caption{Estimated cumulative distribution functions of sensitivity to cost} \label{figure_caseStudy1_cost}
\end{figure}

Figure \ref{figure_caseStudy1} shows the estimated cumulative distribution functions of the implicit attribute values for the DPM-MNL, the two PM-MNL models and the LC-MNL model with 14 mixture components. In addition, Figure \ref{figure_caseStudy1_cost} visualises the estimated cumulative distribution function of the taste parameter capturing sensitivity to cost for the four models. It can be seen that the four models yield distinct representations of taste heterogeneity: By definition, the normal distribution is symmetric and exhibits comparatively light tails. The log-normal only has support on the strictly positive real line and exhibits heavy tails. Both the DPM-MNL and LC-MNL models represent heterogeneity in a discrete fashion so that the estimated cumulative distribution functions are not smooth. 

A closer inspection of the cumulative distribution functions of the implicit attribute values for the the DPM-MNL model reveals several interesting features of the discrete heterogeneity representation produced by the DPM-MNL and LC-MNL model (see Figure \ref{figure_caseStudy1}). First, the heterogeneity distributions under the DPM-MNL and LC-MNL models are not confined to follow any specific parametric function. Therefore, the two models are able to exhibit comparatively light tails in the second orthant without compromising on the flexibility of the heterogeneity representation in the first orthant. By comparison, the normal distribution can assign non-zero amounts of probability mass to the first orthant, but symmetry constraints confine its ability to flexibly represent heterogeneity in the first orthant. Likewise, the strictly positive support of the log-normal distribution comes at the expense of heavy tails in the first orthant. Second, the DPM-MNL and LC-MNL models are capable of endogenously recovering patterns of attribute non-attendance \citep[e.g.][]{hess_its_2013}. For example, in the case of the attributes access time and delay probability, we observe that a small, yet noticeable section of the cumulative distribution functions are almost perfectly aligned with the ordinate, which suggests that a noteworthy proportion of the subjects are insensitive to these attributes.

\FloatBarrier

Furthermore, we observe that the estimated cumulative distribution functions of the random taste parameter capturing sensitivity to cost differ across the considered models (see Figure \ref{figure_caseStudy1_cost}). In the case of the PM-MNL model with normally distributed implicit attribute values, the estimated cumulative distribution function exhibits a rather heavy tail. However, the cumulative distribution functions of the other models do not suggest that such extreme sensitivities are present. We also note that an upper bound of $-0.001$ was imposed on the estimate on cost sensitivity in the DPM-MNL and LC-MNL models to assure finite moments for the distributions of the implicit attribute values \citep[also see][]{daly_assuring_2012}. However, this bound is only relevant for a negligible proportion of the mixture components, and the vast majority of the subjects attend to the cost attribute. 

The differences in in-sample fit and in the shapes of the estimated heterogeneity distributions across the considered models are also reflected in quantitative differences in willingness-to-pay measures. Table \ref{table_caseStudy_wtp} gives summary statistics describing the estimated distributions of the implicit attribute values. In accordance with Figure \ref{figure_caseStudy1}, we observe that the estimated heterogeneity distributions differ noticeably, yet not starkly, in terms of their median values. However, differences in the dispersions of the estimated heterogeneity distributions are more pronounced, as the interquartile ranges (IQRs) and the interdecile ranges (IDRs) differ substantially across the estimated models. Due to the light tails of the normal distributions, the estimated heterogeneity distributions of the PM-MNL model with normal heterogeneity are the least dispersed. By contrast, the estimated heterogeneity distributions of the LC-MNL model with 14 components and the PM-MNL model with log-normal heterogeneity exhibit considerably larger IQRs and IDRs. Due to the influence of the base measure, the marginal heterogeneity distributions of the DPM-MNL model exhibit lighter tails than the LC-MNL model with 14 components and the PM-MNL model with log-normal heterogeneity.

\begin{table}[h]
\small
\input{table_caseStudy_wtp.tex}
\caption{Summary statistics describing the estimated distributions of the implicit attribute values} \label{table_caseStudy_wtp}
\end{table}
\FloatBarrier

\begin{figure}[h]
\centering
\includegraphics[width = \textwidth]{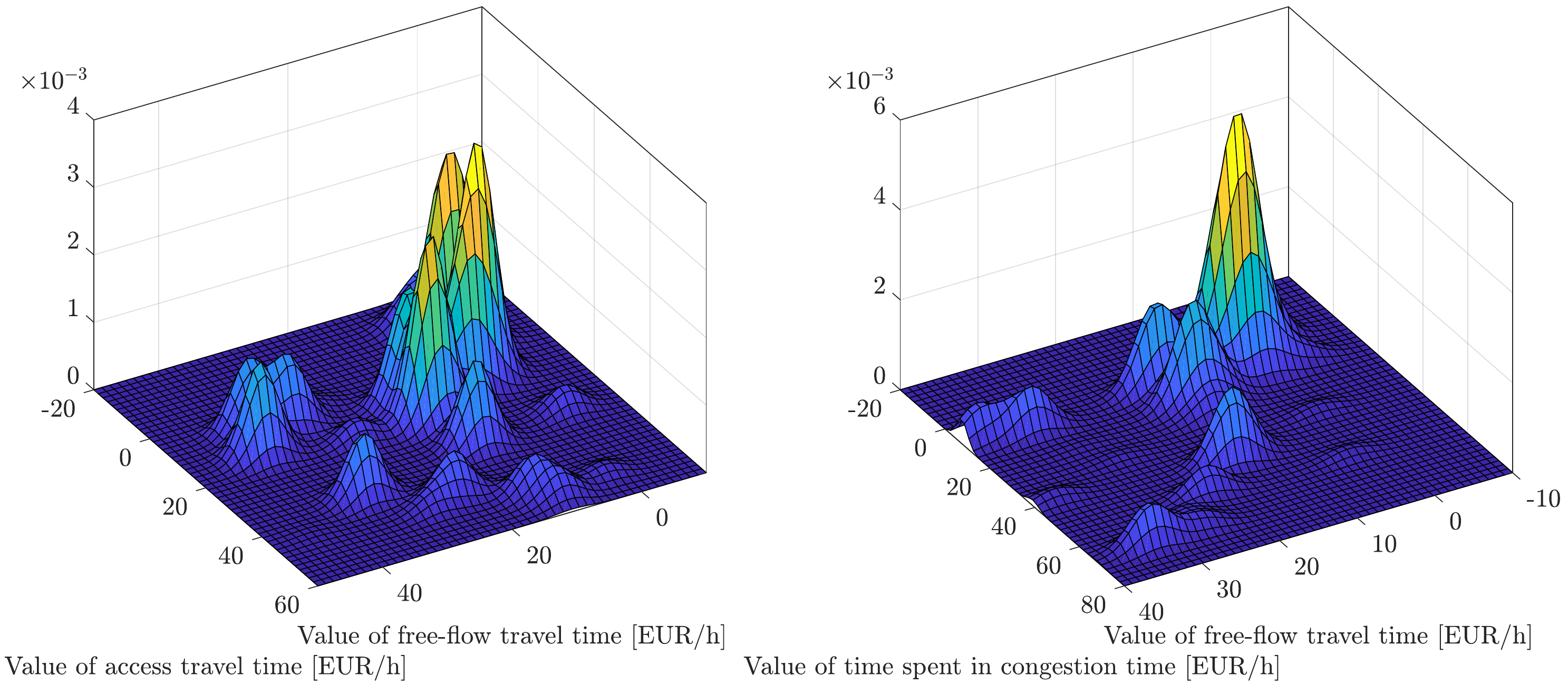}
\caption{Estimated joint densities of the selected pairings of implicit attribute values in the DPM-MNL model} \label{figure_caseStudy1_bi}
\end{figure}

Since the DPM-MNL model also identifies dependencies between random taste parameters, we can meaningfully examine the joint distribution functions of selected pairings of sensitivities and implicit attribute values. Figure \ref{figure_caseStudy1_bi} displays the estimated joint densities of two pairings of implicit attribute values. For both density functions, we observe that the densities are most highly concentrated in regions with low attribute valuations. However, both joint density functions also exhibit outlying peaks, where the valuations of one or both attributes assume comparatively large values.   

For completeness, we report that the estimate of the concentration parameter $\alpha$ is 11.7. Hence, the expected number of mixture components with at least one occupant is 44.

\section{Conclusion} \label{S_conclusion}

The representation of unobserved taste heterogeneity is a key concern of discrete choice analysis. The literature distinguishes two principal approaches to accommodate unobserved taste heterogeneity in M-MNL models, namely PM-MNL and LC-MNL models. Both approaches are widely used in disciplines studying individual choice behaviour but are subject to limitations: PM-MNL models rely on potentially restrictive distributional assumptions, as the shape of the estimated taste parameter distribution is constrained to be equal to the functional form of the imposed parametric random distribution \citep{vij_random_2017}. LC-MNL models free the analyst from parametric assumptions \citep{greene_latent_2003} but require the analyst to exogenously determine the number of mixture components. 

In this paper, we leverage Bayesian nonparametric methods in conjunction with the Dirichlet process to capture unobserved taste heterogeneity in an infinite-dimensional generalisation of the LC-MNL model. Specifically, we present a M-MNL model, which employs the truncated stick-breaking construction of the Dirichlet process as a flexible nonparametric discrete mixing distribution. The resulting model is a Dirichlet process mixture multinomial logit (DPM-MNL) model, which fuses the benefits of a normative mixed random utility model with a flexible statistical method for the representation of unobserved heterogeneity. The DPM-MNL model relies on less restrictive parametric assumptions than a PM-MNL model and in contrast to a LC-MNL model, the analyst is not required to specify the complexity of the discrete mixing distribution a priori, as the DPM-MNL model infers the effective number of non-empty mixture components from the evidence. For posterior inference in the proposed DPM-MNL model, we derive an EM algorithm. In a simulation study, we demonstrate that the proposed model framework can flexibly capture differently-shaped taste parameter distributions. Furthermore, the proposed model framework is empirically validated in a case study on motorists' route choice preferences. We find that in terms of in-sample fit and out-of-sample predictive ability, the proposed DPM-MNL model outperforms a LC-MNL model as well as PM-MNL models with common mixing distributions.

Compared to extant modelling approaches, our proposed DPM-MNL model offers two benefits: First, only limited parametric assumptions about the nature of the base measure are required for the DPM-MNL model to be able to flexibly recover differently-shaped taste parameter distributions. Second, the DPM-MNL model captures heterogeneity distributions nonparametrically, but does not require the analyst to specify the size of the parameter space a priori, as the complexity of the discrete mixing distribution is inferred from the evidence. As a result, the DPM-MNL model can flexibly recover differently-shaped taste parameter distributions and can identify patterns of attribute non-attendance. Overall, the flexible and adaptive nature of the DPM-MNL model framework substantially reduces the need for parametric assumptions and allows for the identification of patterns of heterogeneity, which could otherwise only be revealed through extensive specification searches.

Our proposed model framework is not devoid of limitations: First, the DPM-MNL model yields rather unstructured representations of heterogeneity. Even though it is straightforward to compute summary statistics for the estimated taste parameter distributions (see Table \ref{table_caseStudy_wtp}), certain empirical applications may require smooth representations of taste heterogeneity. Future work could thus explore ways to incorporate semi-nonparametric heterogeneity representations into the proposed model framework by allowing for parametric heterogeneity within the mixture components \citep[see e.g.][]{burda_bayesian_2008,li_bayesian_2013}. A second limitation of the DPM-MNL model concerns the nature of the Dirichlet process. In any Dirichlet process mixture model, the component weights are constructed from independent and identically distributed Beta random variables. Other stochastic processes such as the Pitman-Yor process \citep{pitman_two-parameter_1997} are not subject to this restriction but also exhibit desirable clustering and discreteness properties. Future research may thus explore the use of alternate stochastic processes, which induce random partitions, to accommodate unobserved heterogeneity in discrete choice models. A third limitation of the DPM-MNL model concerns its dependency on the specification of a parametric base measure. In general, the choice of the base measure in Dirichlet process mixture models is opportunistic in that it is driven by computational considerations rather than by behavioural principles \citep{gorur_dirichlet_2010,mcauliffe_nonparametric_2006}. Future research may therefore investigate methods that preclude the a priori specification of a parametric base measure in a DPM-MNL model. For example, \citet{mcauliffe_nonparametric_2006} present an approach to infer the base measure of the Dirichlet process nonparametrically in an empirical Bayes fashion. 

In addition, there are several broader directions for future research to build on the work presented in the current paper: First, it may be worthwhile to apply the proposed model framework to other datasets to assess whether the findings presented in the current paper generalise to other empirical contexts. Second, future research could investigate methods to combine systematic and random representations of heterogeneity \citep[see][]{bhat_endogenous_1997,bhat_incorporating_2000} within a DPM-MNL model. To that end, covariate-dependent stick-breaking processes \citep{ren_logistic_2011,rodriguez_nonparametric_2011} could be leveraged. Ultimately, future research may develop comprehensive simulation-based and empirical comparisons of different methods to accommodate unobserved taste heterogeneity in discrete choice models \citep[see e.g.][]{fosgerau_comparison_2009,keane_comparing_2013}. In particular, it may be worthwhile to contrast the performance of the proposed DPM-MNL model with other nonparametric \citep[e.g.][]{vij_random_2017}, semi-nonparametric \citep[e.g.][]{bastin_estimating_2010,fosgerau_practical_2007,train_em_2008,train_mixed_2016} and parametric \citep[e.g.][]{bhat_new_2012,bhat_new_2017,train_mixed_2005} approaches to account for unobserved heterogeneity in discrete choice models.

Our proposed DPM-MNL model enriches the methodological toolbox of discrete choice analysts by allowing for a flexible, adaptive and parsimonious nonparametric representation of heterogeneity distributions in conjunction with an elegant inference approach. However, there is no one way of catering for taste heterogeneity in discrete choice models, and in any given empirical application, the preferred way of representing taste heterogeneity may depend on a variety of factors such as considerations of parsimony and tractability, emphasis on data fit or predictive performance as well as the analyst's prior belief about the nature of the latent taste parameter distribution.

\newpage
\bibliographystyle{apalike}
\bibliography{2017_imm.bib}

\newpage
\appendix
\section{Appendix}

\subsection{Simulation study}

\begin{table}[h]
\small
\input{table_simStudy02_covariates.tex}
\caption{Data generating process of the simulated covariates for the simulation study} \label{table_simStudy02_covariates}
\end{table}

\begin{table}[h]
\small
\input{table_simStudy02_tastes.tex}
\caption{Data generating processes of the simulated taste parameters for the simulation study} \label{table_simStudy02_tastes}
\end{table}

\end{document}

%% file: model_infinite-mm.tex
\begin{tikzpicture}[x=1.7cm,y=1.8cm]

  % Nodes
  \node[obs]                   (y_nt)      {$y_{n,t}$} ; %
  \node[obs, right=of y_nt, yshift=1.2cm]    (X_nt)      {$\boldsymbol{X}_{n,t}$} ; %
  \node[latent, above=of y_nt,yshift=0.8cm]    (q_n)      {$q_{n}$} ; %
  \node[latent, above=of q_n,yshift=1.2cm]    (pi)      {$\boldsymbol{\pi}$} ; %
  \node[latent, above=of pi,]    (alpha)      {$\alpha$} ; %
  \node[latent, left=of y_nt, yshift=1.2cm]    (beta_k)      {$\boldsymbol{\beta}_{k}$} ; %
  \node[const, above=of y_nt, yshift=-0.3cm]    (k_eq_q)      {$k = q_{n}$} ; %
  \node[const, above=of q_n, yshift=-0.7cm]    (Cat_label)      {\phantom{a}} ; %

  % Factors
  \factor[above=of y_nt]     {y_nt-f}     {MNL} {} {} ; 
  \factor[above=of q_n]     {q_n-f}     {Categorical} {} {} ;  
  \factor[above=of pi]     {pi-f}     {$\mbox{GEM}_{K}$} {} {} ;  
  \factor[above=of beta_k]     {beta_k-f}     {$\mbox{G}_{0}$} {} {} ;  
  \factor[above=of alpha]     {alpha-f}     {$\mbox{f}$} {} {} ;  
  
  % Edges
  \factoredge {X_nt}  {y_nt-f} 	{y_nt} ; %
  \factoredge {alpha}  {pi-f} 	{pi} ; %
  \factoredge {pi}  {q_n-f} 	{q_n} ; %
  \factoredge {beta_k}  {y_nt-f} 	{y_nt} ; %
  \factoredge {}  {beta_k-f} 	{beta_k} ; %
  \factoredge {}  {alpha-f} 	{alpha} ; %
  
  % Gate
  \gate {y_nt-g} {(y_nt-f)(y_nt-f-caption)(k_eq_q)} {q_n}
  
  % Plate
  \plate {plate1} { %
    (y_nt)(y_nt-f)(y_nt-g)%
    (X_nt)%
  } {$t \in \{1,\ldots,T_{n} \}$}; %
  \plate {plate2} {%
    (plate1) %
    (q_n)(q_n-f)(X_nt)(Cat_label)
  }{$n \in \{1, \ldots, N \}$}
  \plate {plate3} {%
    (beta_k)
  }{$k \in \{1,\ldots,K \}$}

\end{tikzpicture}

%% file: model_finite-mm.tex
\begin{tikzpicture}[x=1.7cm,y=1.8cm]

  % Nodes
  \node[obs]                   (y_nt)      {$y_{n,t}$} ; %
  \node[obs, right=of y_nt, yshift=1.2cm]    (X_nt)      {$\boldsymbol{X}_{n,t}$} ; %
  \node[latent, above=of y_nt,yshift=0.8cm]    (q_n)      {$q_{n}$} ; %#
  \node[latent, above=of q_n, yshift=1.2cm]    (pi)      {$\boldsymbol{\pi}$} ; %
  \node[latent, left=of y_nt, yshift=1.2cm]    (beta_k)      {$\boldsymbol{\beta}_{k}$} ; %
  \node[const, above=of y_nt, yshift=-0.3cm]    (k_eq_q)      {$k = q_{n}$} ; %
  \node[const, above=of q_n, yshift=-0.7cm]    (Cat_label)      {\phantom{a}} ; %

  % Factors
  \factor[above=of y_nt]     {y_nt-f}     {MNL} {} {} ; 
  \factor[above=of q_n]     {q_n-f}     {Categorical} {} {} ;   
  
  % Edges
  \factoredge {X_nt}  {y_nt-f} 	{y_nt} ; %
  \factoredge {pi}  {q_n-f} 	{q_n} ; %
  \factoredge {beta_k}  {y_nt-f} 	{y_nt} ; %
  
  % Gate
  \gate {y_nt-g} {(y_nt-f)(y_nt-f-caption)(k_eq_q)} {q_n}
  
  % Plate
  \plate {plate1} { %
    (y_nt)(y_nt-f)(y_nt-g)%
    (X_nt)%
  } {$t \in \{1,\ldots,T_{n} \}$}; %
  \plate {plate2} {%
    (plate1) %
    (q_n)(q_n-f)(X_nt)(Cat_label)
  }{$n \in \{1, \ldots, N \}$}
  \plate {plate3} {%
  (beta_k)
  }{$k \in \{1, \ldots, K \}$}

\end{tikzpicture}

%% file: table_caseStudy_lc_ic.tex
\centering
\ra{1.2}
\begin{tabular}{@{}S[table-format=2.0] S[table-format=3.0] S[table-format=4.1] S[table-format=4.1] S[table-format=4.1] @{}}
\toprule
\multicolumn{1}{c}{\textbf{No. of}}  & \multicolumn{1}{c}{\textbf{No. of}}  & \\
\multicolumn{1}{c}{\textbf{components}}  & \multicolumn{1}{c}{\textbf{parameters}}  & \multicolumn{1}{c}{\textbf{Log-likelihood}}  & \multicolumn{1}{c}{\textbf{AIC}} & \multicolumn{1}{c}{\textbf{BIC}} \\
\midrule
1&	5&	-1709.8&	3429.5&	3450.1\\
2&	11&	-1595.3&	3212.6&	3258.0\\
5&	29&	-1476.6&	3011.1&	3130.6\\
6&	35&	-1457.6&	2985.2&	\textbf{3129.4}\\
7&	41&	-1439.8&	2961.6&	3130.5\\
8&	47&	-1429.1&	2952.2&	3145.8\\
9&	53&	-1422.9&	2951.8&	3170.2\\
10&	59&	-1416.6&	2951.2&	3194.3\\
11&	65&	-1406.1&	2942.1&	3210.0\\
12&	71&	-1398.1&	2938.2&	3230.8\\
13&	77&	-1390.6&	2935.3&	3252.5\\
14&	83&	-1384.2&	\textbf{2934.3}&	3276.3\\
15&	89&	-1383.9&	2945.8&	3312.5\\
\bottomrule
\end{tabular}

%% file: table_caseStudy_comparison.tex
\centering
\ra{1.2}
\begin{tabular}{@{}p{6cm} S[table-format=4.1] S[table-format=3.1] @{}}
\toprule
& \multicolumn{2}{c}{\textbf{Log-likelihood}}  \\ 
\cmidrule{2-3} 
  &   & \multicolumn{1}{c}{\textbf{Out-of-sample fit}} \\
\textbf{Model}  & \multicolumn{1}{c}{\textbf{In-sample fit}}  & \multicolumn{1}{c}{\textbf{(ten-fold cross-validated)}} \\
\midrule
MNL & -1709.8 & -171.9 \\
PM-MNL (normal heterogeneity in implicit attribute values) & -1511.5 & -152.5 \\
PM-MNL (log-normal heterogeneity in implicit attribute values) & -1516.7 & -152.8 \\
LC-MNL ($K = 6$) & -1457.6 & -151.6 \\
LC-MNL ($K = 14$) & -1384.2 & -149.1  \\
DPM-MNL & -1363.1 &  -147.8  \\
\bottomrule
\end{tabular}

%% file: table_caseStudy_wtp.tex
\centering
\ra{1.2}
\begin{tabular}{@{} l S[table-format=3.1] S[table-format=3.1] S[table-format=3.1]  S[table-format=3.1] S[table-format=3.1] c S[table-format=3.1] S[table-format=3.1] @{}}
\toprule
\textbf{Attribute} & \multicolumn{5}{c}{\textbf{Percentiles}} & \phantom{a} & \multicolumn{2}{c}{\textbf{Dispersion}}  \\ 
\cmidrule{2-6} \cmidrule{8-9} 
\quad \textbf{Model}  & 
\multicolumn{1}{c}{\textbf{10\textsuperscript{th}}}  &
\multicolumn{1}{c}{\textbf{25\textsuperscript{th}}}  & 
\multicolumn{1}{c}{\textbf{50\textsuperscript{th}}}  & 
\multicolumn{1}{c}{\textbf{75\textsuperscript{th}}}  & 
\multicolumn{1}{c}{\textbf{90\textsuperscript{th}}}  & & 
\multicolumn{1}{c}{\textbf{IQR}}  & 
\multicolumn{1}{c}{\textbf{IDR}}  \\
\midrule
Free-flow travel time [EUR/h] \\
\quad PM-MNL (normal) & 1.5 & 5.0 & 8.9 & 12.8 & 16.3 & & 7.8 & 14.8 \\
\quad PM-MNL (log-normal) & 1.3 &  3.6 &  10.9 &  33.1 & 89.6 & & 29.5 & 88.2\\
\quad LC-MNL ($K = 14$)  & 1.8 & 4.1 & 9.6 &  25.7 & 57.1 & & 21.6 & 55.3 \\
\quad DPM-MNL & 1.9 &  4.1 &  12.8  &  21.9  &  37.9 & & 17.8 & 35.9 \\
Access time [EUR/h] \\
\quad PM-MNL (normal) & 1.1 &  7.3 &   14.2 &   21.1 &   27.3 & & 13.8 & 26.3 \\
\quad PM-MNL (log-normal) & 2.0  &   4.9 &   13.5 &  37.3 & 93.0 & & 32.4 & 91.0 \\
\quad LC-MNL ($K = 14$) & 1.0 &  7.3 &   20.2  &  30.4  & 60.3 & & 23.1 & 59.4 \\
\quad DPM-MNL & 0.7 & 6.2 &   12.0 & 20.4  & 43.9 & & 14.2 & 43.2 \\
Time spent in congestion [EUR/h] \\
\quad PM-MNL (normal) & 4.2 &   9.1  &  14.6 &   20.1 &   25.1 & & 11.0 & 20.9 \\
\quad PM-MNL (log-normal) & 2.8 &   6.6 &   17.2 &   44.9 &  106.5 & & 38.3 & 103.7 \\
\quad LC-MNL ($K = 14$) & 7.1 & 7.2 &  16.9 &  73.8 &  150.2 & & 66.6 & 143.1 \\
\quad DPM-MNL & 7.1 &    7.6  & 11.3 &  42.1 & 66.9 & & 34.6 & 59.9 \\
Delay probability [EUR/ (100 pp)] \\
\quad PM-MNL (normal) & -5.1 &    0.5 &    6.7 &  12.9 &  18.5 & &12.4 & 23.6 \\
\quad PM-MNL (log-normal) & 0.2 &   0.8   &  3.3 &   14.3 &   53.2 & & 13.5 & 53.0 \\
\quad LC-MNL ($K = 14$) & -0.8 &   1.0 &    5.5  & 36.6 &   37.3 & & 35.6 & 38.2\\
\quad DPM-MNL & -0.6  & 0.7 &  3.2  & 13.6  & 36.5 & & 12.8 & 37.1 \\
\midrule
\multicolumn{9}{l}{{\footnotesize Note: IQR = interquartile range; IDR = interdecile range}} \\
\bottomrule
\end{tabular}

%% file: table_simStudy02_covariates.tex
\centering
\ra{1.2}
\begin{tabular}{@{}l l l l @{}}
\toprule
\textbf{Attribute} & \textbf{Notation} & \textbf{Unit}  & \textbf{Distribution}  \\ 
\midrule
Task-specific attributes \\
\quad Distance  & $s$ & km & $\mbox{U}(2,20)$ \\
Alternative-specific attributes \\
\quad Speed & $v$ & km/h & $\mbox{U}(10,40)$ \\
\quad In-vehicle travel time & $\mbox{ivtt}$ & min & $60 \cdot \frac{s}{v}$\\
\quad Out-of-vehicle travel time & $\mbox{ovtt}$ & min & $\mbox{U}(0,30)$ \\
\quad Cost & $\mbox{cost}$ & \$ & $\mbox{U}(0,2) + \mbox{U}(0,0.7) \cdot s$ \\
\midrule
Note: \\
\multicolumn{4}{l}{U($a$,$b$) denotes a uniform distribution with support over $[a,b]$.} \\
\bottomrule
\end{tabular}

%% file: table_simStudy02_tastes.tex
\centering
\ra{1.2}
\begin{tabular}{@{}l l l l@{}}
\toprule
\textbf{Taste parameter} & \textbf{Notation} & \textbf{Unit}  & \multicolumn{1}{l}{\textbf{Value / distribution}} \\
\midrule
\multicolumn{4}{l}{Experiment I: Uni-modal distribution}\\
\quad Travel time & $(\beta_{\mbox{ivtt}},\beta_{\mbox{ovtt}})$ & \$/h & $\boldsymbol{\mu} = \begin{pmatrix} 10 \\ 15\end{pmatrix}$, $\boldsymbol{\tau} = \begin{pmatrix} 1.5 \\ 2.0 \end{pmatrix}$, $\boldsymbol{\Omega} = \begin{pmatrix} 1 & 0.5 \\ 0.5 & 1 \end{pmatrix}$\\
&&& $\boldsymbol{D}  = \mbox{diag}(\boldsymbol{\tau})$, $\boldsymbol{\Sigma} = \boldsymbol{D} \boldsymbol{\Omega} \boldsymbol{D}$\\
&&& $\begin{pmatrix} \beta_{n,\mbox{ivtt}} & \beta_{n,\mbox{ovtt}} \end{pmatrix}' \sim \mbox{MVN}(\boldsymbol{\mu}, \boldsymbol{\Sigma}), \quad n = 1, \dots, N$ \\
\quad Travel cost & $\beta_{\mbox{cost}}$ & utils/\$ & $\tilde{\beta}_{\mbox{n,cost}} \sim \mbox{N}(0.75, 0.25^{2}), \quad n = 1, \dots, N$ \\ 
&&& $\beta_{n,\mbox{cost}} = -\exp(\tilde{\beta}_{\mbox{n,cost}}), \quad n = 1, \dots, N$ \\

\multicolumn{4}{l}{Experiment II: Uni-modal distribution with one heavy-tailed marginal}\\
\quad Travel time & $(\beta_{\mbox{ivtt}},\beta_{\mbox{ovtt}})$ & \$/h & $\boldsymbol{\mu} = \begin{pmatrix} 12.0 \\ 2.8\end{pmatrix}$, $\boldsymbol{\tau} = \begin{pmatrix} 1.5 \\ 0.3 \end{pmatrix}$, $\boldsymbol{\Omega} = \begin{pmatrix} 1 & 0.3 \\ 0.3 & 1 \end{pmatrix}$\\
&&& $\boldsymbol{D}  = \mbox{diag}(\boldsymbol{\tau})$, $\boldsymbol{\Sigma} = \boldsymbol{D} \boldsymbol{\Omega} \boldsymbol{D}$\\
&&& $\begin{pmatrix} \beta_{n,\mbox{ivtt}} & \beta_{n,\mbox{ovtt}} \end{pmatrix}' \sim \mbox{MVN}(\boldsymbol{\mu}, \boldsymbol{\Sigma}), \quad n = 1, \dots, N$ \\
&&& $\beta_{n,\mbox{ovtt}} = \exp(\tilde{\beta}_{\mbox{n,ovtt}}), \quad n = 1, \dots, N$ \\
\quad Travel cost & $\beta_{\mbox{cost}}$ & utils/\$ & $\tilde{\beta}_{\mbox{n,cost}} \sim \mbox{N}(0.60, 0.25^{2}), \quad n = 1, \dots, N$ \\ 
&&& $\beta_{n,\mbox{cost}} = -\exp(\tilde{\beta}_{\mbox{n,cost}}), \quad n = 1, \dots, N$ \\

\multicolumn{4}{l}{Experiment III: Bi-modal distribution}\\
\quad Travel time & $(\beta_{\mbox{ivtt}},\beta_{\mbox{ovtt}})$ & \$/h & $\boldsymbol{\mu}_{1} = \begin{pmatrix} 12 \\ 16\end{pmatrix}$, $\boldsymbol{\tau}_{1} = \begin{pmatrix} 1 \\ 2 \end{pmatrix}$, $\boldsymbol{\Omega}_{1} = \begin{pmatrix} 1 & 0.2 \\ 0.2 & 1 \end{pmatrix}$\\
&&& $\boldsymbol{\mu}_{2} = \begin{pmatrix} 6 \\ 10\end{pmatrix}$, $\boldsymbol{\tau}_{2} = \begin{pmatrix} 1 \\ 2 \end{pmatrix}$, $\boldsymbol{\Omega}_{2} = \begin{pmatrix} 1 & -0.4 \\ -0.4 & 1 \end{pmatrix}$\\
&&& $\boldsymbol{D}_{k}  = \mbox{diag}(\boldsymbol{\tau}_{k})$, $\boldsymbol{\Sigma}_{k} = \boldsymbol{D}_{k} \boldsymbol{\Omega}_{k} \boldsymbol{D}_{k}, \quad k = 1, \dots, K$ \\
&&& $q_{n} \sim \mbox{Categorical} \left ( \begin{pmatrix} 0.75 & 0.25 \end{pmatrix}' \right ), \quad n = 1, \dots, N$ \\
&&& $\begin{pmatrix} \beta_{n,\mbox{ivtt}} & \beta_{n,\mbox{ovtt}} \end{pmatrix}' \sim \mbox{MVN}(\boldsymbol{\mu}_{q_{n}}, \boldsymbol{\Sigma}_{q_{n}}), \quad n = 1, \dots, N$ \\
\quad Travel cost & $\beta_{\mbox{cost}}$ & utils/\$ & $\tilde{\beta}_{\mbox{n,cost}} \sim \mbox{N}(0.80, 0.25^{2}), \quad n = 1, \dots, N$ \\ 
&&& $\beta_{n,\mbox{cost}} = -\exp(\tilde{\beta}_{\mbox{n,cost}}), \quad n = 1, \dots, N$ \\

\multicolumn{4}{l}{Experiment IV: Tri-modal distribution}\\
\quad Travel time & $(\beta_{\mbox{ivtt}},\beta_{\mbox{ovtt}})$ & \$/h & $\boldsymbol{\mu}_{1} = \begin{pmatrix} 10.00 \\ 15.00 \end{pmatrix}$, $\boldsymbol{\Sigma}_{1} = \begin{pmatrix} 2^{2} & 0 \\ 0 & 2^{2} \end{pmatrix}$\\
&&& $\boldsymbol{\mu}_{2} = \begin{pmatrix} 0.88 \\ 24.12 \end{pmatrix}$, $\boldsymbol{\Sigma}_{2} = \begin{pmatrix} 1.2^{2} & 0 \\ 0 & 1.2^{2} \end{pmatrix}$\\
&&& $\boldsymbol{\mu}_{3} = \begin{pmatrix} 19.12 \\ 24.12 \end{pmatrix}$, $\boldsymbol{\Sigma}_{3} = \begin{pmatrix} 1.8^{2} & 0 \\ 0 & 1.2^{2} \end{pmatrix}$\\
&&& $q_{n} \sim \mbox{Categorical} \left ( \begin{pmatrix} 0.35 & 0.25 & 0.40 \end{pmatrix}' \right ), \quad n = 1, \dots, N$ \\
&&& $\begin{pmatrix} \beta_{n,\mbox{ivtt}} & \beta_{n,\mbox{ovtt}} \end{pmatrix}' \sim \mbox{MVN}(\boldsymbol{\mu}_{q_{n}}, \boldsymbol{\Sigma}_{q_{n}}), \quad n = 1, \dots, N$ \\
\quad Travel cost & $\beta_{\mbox{cost}}$ & utils/\$ & $\tilde{\beta}_{\mbox{n,cost}} \sim \mbox{N}(0.60, 0.25^{2}), \quad n = 1, \dots, N$ \\ 
&&& $\beta_{n,\mbox{cost}} = -\exp(\tilde{\beta}_{\mbox{n,cost}}), \quad n = 1, \dots, N$ \\

\midrule
Note: \\
\multicolumn{4}{l}{$\mbox{N}(a,b^{2})$ denotes a univariate normal distribution with location $a$ and scale $b$.} \\
\multicolumn{4}{l}{$\mbox{MVN}(\boldsymbol{\mu}, \boldsymbol{\Sigma})$ denotes a multivariate normal distribution with mean vector $\boldsymbol{\mu}$ and covariance $\boldsymbol{\Sigma}$.} \\
\multicolumn{4}{l}{$\boldsymbol{\Sigma} = \boldsymbol{D} \boldsymbol{\Omega} \boldsymbol{D}$, $\boldsymbol{D}  = \mbox{diag}(\boldsymbol{\tau})$ with correlation matrix  $\boldsymbol{\Omega}$ and scale vector $\boldsymbol{\tau}$.} \\
\bottomrule
\end{tabular}

%% file: manuscript_stick-breaking-v19_rk_17.01.2018-arxiv.bbl
\begin{thebibliography}{}

\bibitem[Akaike, 1974]{akaike_new_1974}
Akaike, H. (1974).
\newblock A new look at the statistical model identification.
\newblock {\em IEEE Transactions on Automatic Control}, 19(6):716--723.

\bibitem[Aldous, 1985]{aldous_exchangeability_1985}
Aldous, D.~J. (1985).
\newblock Exchangeability and related topics.
\newblock In {\em École d'Été de {Probabilités} de {Saint}-{Flour} {XIII}
  — 1983}, pages 1--198. Springer, Berlin, Heidelberg.
\newblock DOI: 10.1007/BFb0099421.

\bibitem[Ansari and Iyengar, 2006]{ansari_semiparametric_2006}
Ansari, A. and Iyengar, R. (2006).
\newblock Semiparametric {Thurstonian} {Models} for {Recurrent} {Choices}: {A}
  {Bayesian} {Analysis}.
\newblock {\em Psychometrika}, 71(4):631.

\bibitem[Antoniak, 1974]{antoniak_mixtures_1974}
Antoniak, C.~E. (1974).
\newblock Mixtures of {Dirichlet} {Processes} with {Applications} to {Bayesian}
  {Nonparametric} {Problems}.
\newblock {\em The Annals of Statistics}, 2(6):1152--1174.

\bibitem[Axhausen et~al., 2015]{axhausen_ermittlung_2015}
Axhausen, K.~W., Ehreke, I., Glemser, A., Hess, S., Jödden, C., Nagel, K.,
  Sauer, A., and Weis, C. (2015).
\newblock Ermittlung von {Bewertungsansätzen} für {Reisezeiten} und
  {Zuverlässigkeit} auf der {Basis} eines {Modells} für modale
  {Verlagerungen} im nicht-gewerblichen und gewerblichen {Personenverkehr} für
  die {Bundesverkehrswegeplanung}.

\bibitem[Bastin et~al., 2010]{bastin_estimating_2010}
Bastin, F., Cirillo, C., and Toint, P.~L. (2010).
\newblock Estimating {Nonparametric} {Random} {Utility} {Models} with an
  {Application} to the {Value} of {Time} in {Heterogeneous} {Populations}.
\newblock {\em Transportation Science}, 44(4):537--549.

\bibitem[Bhat, 1997]{bhat_endogenous_1997}
Bhat, C.~R. (1997).
\newblock An {Endogenous} {Segmentation} {Mode} {Choice} {Model} with an
  {Application} to {Intercity} {Travel}.
\newblock {\em Transportation Science}, 31(1):34--48.

\bibitem[Bhat, 1998]{bhat_accommodating_1998}
Bhat, C.~R. (1998).
\newblock Accommodating variations in responsiveness to level-of-service
  measures in travel mode choice modeling.
\newblock {\em Transportation Research Part A: Policy and Practice},
  32(7):495--507.

\bibitem[Bhat, 2000]{bhat_incorporating_2000}
Bhat, C.~R. (2000).
\newblock Incorporating {Observed} and {Unobserved} {Heterogeneity} in {Urban}
  {Work} {Travel} {Mode} {Choice} {Modeling}.
\newblock {\em Transportation Science}, 34(2):228--238.

\bibitem[Bhat and Lavieri, 2017]{bhat_new_2017}
Bhat, C.~R. and Lavieri, P.~S. (2017).
\newblock A new mixed {MNP} model accommodating a variety of dependent
  non-normal coefficient distributions.
\newblock {\em Theory and Decision}, pages 1--37.

\bibitem[Bhat and Sidharthan, 2012]{bhat_new_2012}
Bhat, C.~R. and Sidharthan, R. (2012).
\newblock A new approach to specify and estimate non-normally mixed multinomial
  probit models.
\newblock {\em Transportation Research Part B: Methodological}, 46(7):817--833.

\bibitem[Bierlaire, 2016]{bierlaire_pythonbiogeme:_2016}
Bierlaire, M. (2016).
\newblock {PythonBiogeme}: a short introduction.

\bibitem[Blackwell and MacQueen, 1973]{blackwell_ferguson_1973}
Blackwell, D. and MacQueen, J.~B. (1973).
\newblock Ferguson {Distributions} {Via} {Polya} {Urn} {Schemes}.
\newblock {\em The Annals of Statistics}, 1(2):353--355.

\bibitem[Blei and Jordan, 2006]{blei_variational_2006}
Blei, D.~M. and Jordan, M.~I. (2006).
\newblock Variational inference for {Dirichlet} process mixtures.
\newblock {\em Bayesian Analysis}, 1(1):121--143.

\bibitem[Blei et~al., 2017]{blei_variational_2017}
Blei, D.~M., Kucukelbir, A., and McAuliffe, J.~D. (2017).
\newblock Variational {Inference}: {A} {Review} for {Statisticians}.
\newblock {\em Journal of the American Statistical Association},
  112(518):859--877.

\bibitem[Braun and McAuliffe, 2010]{braun_variational_2010}
Braun, M. and McAuliffe, J. (2010).
\newblock Variational {Inference} for {Large}-{Scale} {Models} of {Discrete}
  {Choice}.
\newblock {\em Journal of the American Statistical Association},
  105(489):324--335.

\bibitem[Bujosa et~al., 2010]{bujosa_combining_2010}
Bujosa, A., Riera, A., and Hicks, R.~L. (2010).
\newblock Combining {Discrete} and {Continuous} {Representations} of
  {Preference} {Heterogeneity}: {A} {Latent} {Class} {Approach}.
\newblock {\em Environmental and Resource Economics}, 47(4):477--493.

\bibitem[Burda et~al., 2008]{burda_bayesian_2008}
Burda, M., Harding, M., and Hausman, J. (2008).
\newblock A {Bayesian} mixed logit–probit model for multinomial choice.
\newblock {\em Journal of Econometrics}, 147(2):232--246.

\bibitem[Carvalho et~al., 2010]{carvalho_particle_2010}
Carvalho, C.~M., Lopes, H.~F., Polson, N.~G., and Taddy, M.~A. (2010).
\newblock Particle learning for general mixtures.
\newblock {\em Bayesian Analysis}, 5(4):709--740.

\bibitem[Connor and Mosimann, 1969]{connor_concepts_1969}
Connor, R.~J. and Mosimann, J.~E. (1969).
\newblock Concepts of {Independence} for {Proportions} with a {Generalization}
  of the {Dirichlet} {Distribution}.
\newblock {\em Journal of the American Statistical Association},
  64(325):194--206.

\bibitem[Daly et~al., 2012]{daly_assuring_2012}
Daly, A., Hess, S., and Train, K. (2012).
\newblock Assuring finite moments for willingness to pay in random coefficient
  models.
\newblock {\em Transportation}, 39(1):19--31.

\bibitem[Dempster et~al., 1977]{dempster_maximum_1977}
Dempster, A.~P., Laird, N.~M., and Rubin, D.~B. (1977).
\newblock Maximum {Likelihood} from {Incomplete} {Data} via the {EM}
  {Algorithm}.
\newblock {\em Journal of the Royal Statistical Society. Series B
  (Methodological)}, 39(1):1--38.

\bibitem[Dong and Koppelman, 2014]{dong_comparison_2014}
Dong, X. and Koppelman, F.~S. (2014).
\newblock Comparison of continuous and discrete representations of unobserved
  heterogeneity in logit models.
\newblock {\em Journal of Marketing Analytics}, 2(1):43--58.

\bibitem[Ehreke et~al., 2014]{ehreke_experiences_2014}
Ehreke, I., Hess, S., and Axhausen, K.~W. (2014).
\newblock Experiences from the {German} value of time ({VOT}) and value of
  reliability ({VOR}) study.
\newblock In {\em 10th {International} {Conference} on {Transport} {Survey}
  {Methods} ({ISCTSC}14), {Leura}}.

\bibitem[Ferguson, 1973]{ferguson_bayesian_1973}
Ferguson, T.~S. (1973).
\newblock A {Bayesian} {Analysis} of {Some} {Nonparametric} {Problems}.
\newblock {\em The Annals of Statistics}, 1(2):209--230.

\bibitem[Fosgerau and Bierlaire, 2007]{fosgerau_practical_2007}
Fosgerau, M. and Bierlaire, M. (2007).
\newblock A practical test for the choice of mixing distribution in discrete
  choice models.
\newblock {\em Transportation Research Part B: Methodological}, 41(7):784--794.

\bibitem[Fosgerau and Hess, 2009]{fosgerau_comparison_2009}
Fosgerau, M. and Hess, S. (2009).
\newblock A comparison of methods for representing random taste heterogeneity
  in discrete choice models.
\newblock {\em European Transport-Trasporti Europei}, 42:1--25.

\bibitem[Gelman et~al., 2013]{gelman_bayesian_2013}
Gelman, A., Carlin, J.~B., Stern, H.~S., Dunson, D.~B., Vehtari, A., and Rubin,
  D.~B. (2013).
\newblock {\em Bayesian {Data} {Analysis}, {Third} {Edition}}.
\newblock CRC Press.

\bibitem[Gershman and Blei, 2012]{gershman_tutorial_2012}
Gershman, S.~J. and Blei, D.~M. (2012).
\newblock A tutorial on {Bayesian} nonparametric models.
\newblock {\em Journal of Mathematical Psychology}, 56(1):1--12.

\bibitem[Greene and Hensher, 2003]{greene_latent_2003}
Greene, W.~H. and Hensher, D.~A. (2003).
\newblock A latent class model for discrete choice analysis: contrasts with
  mixed logit.
\newblock {\em Transportation Research Part B: Methodological}, 37(8):681--698.

\bibitem[Greene and Hensher, 2013]{greene_revealing_2013}
Greene, W.~H. and Hensher, D.~A. (2013).
\newblock Revealing additional dimensions of preference heterogeneity in a
  latent class mixed multinomial logit model.
\newblock {\em Applied Economics}, 45(14):1897--1902.

\bibitem[Görür and Rasmussen, 2010]{gorur_dirichlet_2010}
Görür, D. and Rasmussen, C.~E. (2010).
\newblock Dirichlet {Process} {Gaussian} {Mixture} {Models}: {Choice} of the
  {Base} {Distribution}.
\newblock {\em Journal of Computer Science and Technology}, 25(4):653--664.

\bibitem[Hess et~al., 2005]{hess_estimation_2005}
Hess, S., Bierlaire, M., and Polak, J.~W. (2005).
\newblock Estimation of value of travel-time savings using mixed logit models.
\newblock {\em Transportation Research Part A: Policy and Practice},
  39(2):221--236.

\bibitem[Hess et~al., 2013]{hess_its_2013}
Hess, S., Stathopoulos, A., Campbell, D., O’Neill, V., and Caussade, S.
  (2013).
\newblock It’s not that {I} don’t care, {I} just don’t care very much:
  confounding between attribute non-attendance and taste heterogeneity.
\newblock {\em Transportation}, 40(3):583--607.

\bibitem[Hess et~al., 2006]{hess_use_2006}
Hess, S., Train, K.~E., and Polak, J.~W. (2006).
\newblock On the use of a {Modified} {Latin} {Hypercube} {Sampling} ({MLHS})
  method in the estimation of a {Mixed} {Logit} {Model} for vehicle choice.
\newblock {\em Transportation Research Part B: Methodological}, 40(2):147--163.

\bibitem[Ishwaran and James, 2001]{ishwaran_gibbs_2001}
Ishwaran, H. and James, L.~F. (2001).
\newblock Gibbs {Sampling} {Methods} for {Stick}-{Breaking} {Priors}.
\newblock {\em Journal of the American Statistical Association},
  96(453):161--173.

\bibitem[Ishwaran and James, 2002]{ishwaran_approximate_2002}
Ishwaran, H. and James, L.~F. (2002).
\newblock Approximate {Dirichlet} {Process} {Computing} in {Finite} {Normal}
  {Mixtures}: {Smoothing} and {Prior} {Information}.
\newblock {\em Journal of Computational and Graphical Statistics},
  11(3):508--532.

\bibitem[Ishwaran and Zarepour, 2000]{ishwaran_markov_2000}
Ishwaran, H. and Zarepour, M. (2000).
\newblock Markov {Chain} {Monte} {Carlo} in {Approximate} {Dirichlet} and
  {Beta} {Two}-{Parameter} {Process} {Hierarchical} {Models}.
\newblock {\em Biometrika}, 87(2):371--390.

\bibitem[Kamakura and Russell, 1989]{kamakura_probabilistic_1989}
Kamakura, W.~A. and Russell, G.~J. (1989).
\newblock A {Probabilistic} {Choice} {Model} for {Market} {Segmentation} and
  {Elasticity} {Structure}.
\newblock {\em Journal of Marketing Research}, 26(4):379--390.

\bibitem[Keane and Wasi, 2013]{keane_comparing_2013}
Keane, M. and Wasi, N. (2013).
\newblock Comparing {Alternative} {Models} of {Heterogeneity} in {Consumer}
  {Choice} {Behavior}.
\newblock {\em Journal of Applied Econometrics}, 28(6):1018--1045.

\bibitem[Kim et~al., 2004]{kim_assessing_2004}
Kim, J.~G., Menzefricke, U., and Feinberg, F.~M. (2004).
\newblock Assessing {Heterogeneity} in {Discrete} {Choice} {Models} {Using} a
  {Dirichlet} {Process} {Prior}.
\newblock {\em Review of Marketing Science}, 2(1).

\bibitem[Li and Ansari, 2013]{li_bayesian_2013}
Li, Y. and Ansari, A. (2013).
\newblock A {Bayesian} {Semiparametric} {Approach} for {Endogeneity} and
  {Heterogeneity} in {Choice} {Models}.
\newblock {\em Management Science}, 60(5):1161--1179.

\bibitem[McAuliffe et~al., 2006]{mcauliffe_nonparametric_2006}
McAuliffe, J.~D., Blei, D.~M., and Jordan, M.~I. (2006).
\newblock Nonparametric empirical {Bayes} for the {Dirichlet} process mixture
  model.
\newblock {\em Statistics and Computing}, 16(1):5--14.

\bibitem[McFadden and Train, 2000]{mcfadden_mixed_2000}
McFadden, D. and Train, K. (2000).
\newblock Mixed {MNL} models for discrete response.
\newblock {\em Journal of Applied Econometrics}, 15(5):447--470.

\bibitem[McLachlan and Krishnan, 2008]{mclachlan_em_2008}
McLachlan, G.~J. and Krishnan, T. (2008).
\newblock {\em The {EM} algorithm and extensions}.
\newblock Wiley-Interscience.

\bibitem[Neal, 2000]{neal_markov_2000}
Neal, R.~M. (2000).
\newblock Markov {Chain} {Sampling} {Methods} for {Dirichlet} {Process}
  {Mixture} {Models}.
\newblock {\em Journal of Computational and Graphical Statistics},
  9(2):249--265.

\bibitem[Ohlssen et~al., 2007]{ohlssen_flexible_2007}
Ohlssen, D.~I., Sharples, L.~D., and Spiegelhalter, D.~J. (2007).
\newblock Flexible random-effects models using {Bayesian} semi-parametric
  models: applications to institutional comparisons.
\newblock {\em Statistics in Medicine}, 26(9):2088--2112.

\bibitem[Orbanz and Teh, 2011]{orbanz_bayesian_2011}
Orbanz, P. and Teh, Y.~W. (2011).
\newblock Bayesian {Nonparametric} {Models}.
\newblock In Sammut, C. and Webb, G.~I., editors, {\em Encyclopedia of
  {Machine} {Learning}}, pages 81--89. Springer US.
\newblock DOI: 10.1007/978-0-387-30164-8\_66.

\bibitem[Pitman, 2006]{pitman_sequential_2006}
Pitman, J. (2006).
\newblock Sequential constructions of random partitions.
\newblock In Picard, J., editor, {\em Combinatorial {Stochastic} {Processes}},
  number 1875 in Lecture {Notes} in {Mathematics}, pages 55--75. Springer
  Berlin Heidelberg.
\newblock DOI: 10.1007/3-540-34266-4\_4.

\bibitem[Pitman and Yor, 1997]{pitman_two-parameter_1997}
Pitman, J. and Yor, M. (1997).
\newblock The two-parameter {Poisson}-{Dirichlet} distribution derived from a
  stable subordinator.
\newblock {\em The Annals of Probability}, 25(2):855--900.

\bibitem[Rasmussen, 1999]{rasmussen_infinite_1999}
Rasmussen, C.~E. (1999).
\newblock The infinite {Gaussian} mixture model.
\newblock In {\em {NIPS}}, volume~12, pages 554--560.

\bibitem[Ren et~al., 2011]{ren_logistic_2011}
Ren, L., Du, L., Carin, L., and Dunson, D. (2011).
\newblock Logistic {Stick}-{Breaking} {Process}.
\newblock {\em Journal of Machine Learning Research}, 12(Jan):203--239.

\bibitem[Revelt and Train, 1998]{revelt_mixed_1998}
Revelt, D. and Train, K. (1998).
\newblock Mixed {Logit} with {Repeated} {Choices}: {Households}' {Choices} of
  {Appliance} {Efficiency} {Level}.
\newblock {\em The Review of Economics and Statistics}, 80(4):647--657.

\bibitem[Rodríguez and Dunson, 2011]{rodriguez_nonparametric_2011}
Rodríguez, A. and Dunson, D.~B. (2011).
\newblock Nonparametric {Bayesian} models through probit stick-breaking
  processes.
\newblock {\em Bayesian Analysis}, 6(1):145--177.

\bibitem[Schwarz, 1978]{schwarz_estimating_1978}
Schwarz, G. (1978).
\newblock Estimating the {Dimension} of a {Model}.
\newblock {\em The Annals of Statistics}, 6(2):461--464.

\bibitem[Sethuraman, 1994]{sethuraman_constructive_1994}
Sethuraman, J. (1994).
\newblock A {Constructive} {Definition} of {Dirichlet} {Priors}.
\newblock {\em Statistica Sinica}, 4(2):639--650.

\bibitem[Sohn, 2016]{sohn_expectation-maximization_2016}
Sohn, K. (2016).
\newblock An {Expectation}-{Maximization} {Algorithm} to {Estimate} the
  {Integrated} {Choice} and {Latent} {Variable} {Model}.
\newblock {\em Transportation Science}, 51(3):946--967.

\bibitem[Teh, 2011]{teh_dirichlet_2011}
Teh, Y.~W. (2011).
\newblock Dirichlet {Process}.
\newblock In Sammut, C. and Webb, G.~I., editors, {\em Encyclopedia of
  {Machine} {Learning}}, pages 280--287. Springer US.
\newblock DOI: 10.1007/978-0-387-30164-8\_219.

\bibitem[Train, 2016]{train_mixed_2016}
Train, K. (2016).
\newblock Mixed logit with a flexible mixing distribution.
\newblock {\em Journal of Choice Modelling}, 19:40--53.

\bibitem[Train and Sonnier, 2005]{train_mixed_2005}
Train, K. and Sonnier, G. (2005).
\newblock Mixed {Logit} with {Bounded} {Distributions} of {Correlated}
  {Partworths}.
\newblock In {\em Applications of {Simulation} {Methods} in {Environmental} and
  {Resource} {Economics}}, The {Economics} of {Non}-{Market} {Goods} and
  {Resources}, pages 117--134. Springer, Dordrecht.
\newblock DOI: 10.1007/1-4020-3684-1\_7.

\bibitem[Train, 2008]{train_em_2008}
Train, K.~E. (2008).
\newblock {EM} {Algorithms} for nonparametric estimation of mixing
  distributions.
\newblock {\em Journal of Choice Modelling}, 1(1):40--69.

\bibitem[Train, 2009]{train_discrete_2009}
Train, K.~E. (2009).
\newblock {\em Discrete {Choice} {Methods} with {Simulation}}.
\newblock Cambridge University Press, 2nd edition.

\bibitem[Vij and Krueger, 2017]{vij_random_2017}
Vij, A. and Krueger, R. (2017).
\newblock Random taste heterogeneity in discrete choice models: {Flexible}
  nonparametric finite mixture distributions.
\newblock {\em Transportation Research Part B: Methodological}, 106(Supplement
  C):76--101.

\bibitem[Wainwright and Jordan, 2008]{wainwright_graphical_2008}
Wainwright, M.~J. and Jordan, M.~I. (2008).
\newblock Graphical {Models}, {Exponential} {Families}, and {Variational}
  {Inference}.
\newblock {\em Foundations and Trends® in Machine Learning}, 1(1-2):1--305.

\bibitem[Walker and Ben-Akiva, 2002]{walker_generalized_2002}
Walker, J. and Ben-Akiva, M. (2002).
\newblock Generalized random utility model.
\newblock {\em Mathematical Social Sciences}, 43(3):303--343.

\bibitem[Wang and Dunson, 2011]{wang_fast_2011}
Wang, L. and Dunson, D.~B. (2011).
\newblock Fast {Bayesian} {Inference} in {Dirichlet} {Process} {Mixture}
  {Models}.
\newblock {\em Journal of Computational and Graphical Statistics},
  20(1):196--216.

\bibitem[Wedel et~al., 1999]{wedel_discrete_1999}
Wedel, M., Kamakura, W., Arora, N., Bemmaor, A., Chiang, J., Elrod, T.,
  Johnson, R., Lenk, P., Neslin, S., and Poulsen, C.~S. (1999).
\newblock Discrete and {Continuous} {Representations} of {Unobserved}
  {Heterogeneity} in {Choice} {Modeling}.
\newblock {\em Marketing Letters}, 10(3):219--232.

\bibitem[Yuan et~al., 2015]{yuan_guide_2015}
Yuan, Y., You, W., and Boyle, K.~J. (2015).
\newblock A guide to heterogeneity features captured by parametric and
  nonparametric mixing distributions for the mixed logit model.
\newblock In {\em 2015 {AAEA} \& {WAEA} {Joint} {Annual} {Meeting}, {July}
  26-28, {San} {Francisco}, {California}}.

\bibitem[Zhou and Lange, 2010]{zhou_mm_2010}
Zhou, H. and Lange, K. (2010).
\newblock {MM} {Algorithms} for {Some} {Discrete} {Multivariate}
  {Distributions}.
\newblock {\em Journal of Computational and Graphical Statistics},
  19(3):645--665.

\end{thebibliography}
